\documentclass[journal]{IEEEtran}
\usepackage{amsmath,amsfonts,amssymb}
\usepackage{algorithmic}
\usepackage{algorithm}
\usepackage{array}
\usepackage[caption=false,font=scriptsize,labelfont=rm,textfont=rm]{subfig}
\usepackage{textcomp}
\usepackage{stfloats}
\usepackage{url}
\usepackage{verbatim}
\usepackage{graphicx}
\usepackage{cite}
\usepackage{makecell}
\usepackage{xcolor}  
\usepackage{tabularray}
\usepackage{multirow}
\usepackage{soul}
\usepackage{fancyhdr}
\usepackage[figuresright]{rotating}
\begin{document}
\title{An Accurate Beam-Tracking Algorithm with Adaptive Beam Reconstruction via UAV-BSs for Mobile Users}
\author{
Jing Zhang, \IEEEmembership{Member,~IEEE},
Sheng Gao,
Xin Feng,
Hongwei Yang, and
Geng Sun,  \IEEEmembership{Member,~IEEE}
\thanks{This research was supported in part by the  Jilin Province Science and Technology Development Plan Project (YDZJ202201ZYTS416), and in part by the National Natural Science
Foundation of China (62272194). (\emph{Corresponding author: Geng Sun})}
\thanks{Jing Zhang is with the College of Computer Science and Technology, Changchun University of Science and Technology, Changchun 130022, China, and also with the College of Communication Engineering, Jilin University, Changchun 130012, China (e-mail: zhang\_jing@cust.edu.cn).}
\thanks{Sheng Gao, Xin Feng and Hongwei Yang are with the College of Computer Science and Technology, Changchun University of Science and Technology, Changchun 130022, China (e-mail: shenggao.cs@gmail.com; fengxin@cust.edu.cn; yanghongwei@cust.edu.cn).}
\thanks{Geng Sun is with the College of Computer Science and Technology, Jilin University, Changchun 130012, China, and also with the College of Computing and Data Science, Nanyang Technological University, Singapore 639798 (e-mail: sungeng@jlu.edu.cn).}}
\markboth{Journal of \LaTeX\ Class Files,~Vol.~14, No.~8, August~2021}%
{Shell \MakeLowercase{\textit{et al.}}: A Sample Article Using IEEEtran.cls for IEEE Journals}


\maketitle
\IEEEpubid{\begin{minipage}{\textwidth}\ \\[40pt] \centering
		Copyright (c) 20xx IEEE. Personal use of this material is permitted. However, permission to use this material for any other purposes \\ must be obtained from the IEEE by sending a request to pubs-permissions@ieee.org.
\end{minipage}}
\begin{abstract}
Unmanned aerial vehicles (UAVs) with flexible deployment contribute to enlarging the distance of information transmission to mobile users (MUs) in constrained environment. However, due to the high mobility of both UAVs and MUs, it is challenging to establish an accurate beam towards the target MU with high beam gain in real-time. In this study, UAV base stations (UAV-BSs) consisting of position-known assisted UAVs (A-UAVs) and position-unknown assisted UAVs (U-UAVs) are employed to transmit data to MUs. Specifically, a bi-directional angle-aware beam tracking with adaptive beam reconstruction (BAB-AR) algorithm is proposed to construct an optimal beam that can quickly adapt the movement of the target MU. First, the angle-aware beam tracking is realized within the UAV-BSs using a proposed global dynamic crow search algorithm without historical trajectory. Furthermore, the Gaussian process regression model is trained by A-UAVs to predict the azimuth and elevation angles of MUs. Meanwhile, we focus on the beam width and design a time interval adjustment mechanism for adaptive beam reconstruction to track high-speed MUs. Finally, the performance of the BAB-AR algorithm is compared with that of benchmark algorithms, and simulate results verifies that the BAB-AR algorithm can construct an accurate beam capable of covering high-speed MUs with the half power beam width in a timely manner.
\end{abstract}


\section{Introduction}
\IEEEPARstart{B}{eam-tracking} is a promising technique for real-time transmission of data to mobile users (MUs), offering a high signal-to-noise ratio (SNR) with millimeter wave (mmWave) \cite{JHYu2021,JWZhao2021}. For instance, the beam gain for a MU can be increased significantly by quickly reconstructing the beam \cite{YJKim2018,SJayaprakasam2017}. In vehicular communication systems, beam tracking enables the transmission of data between the base station (BS) and target MU at high data rate and with dependable service \cite{JSMu2021}. However, in remote scenarios such as mountainous areas, the beam-tracking service provided by a BS may be interrupted by tall buildings or plants\cite{JTan2020,TXFeng2022}. 

\par In recent years, UAV-BSs have found widespread application in various domains, including civilian and military contexts, to ensure reliable communications for MUs without any environmental constraints \cite{JHLi2023,ZWang2023}. Due to the mobility of UAVs, UAV-BSs can be flexibly deployed to establish a line-of-sight (LoS) communication link \cite{TTaleb2021,BLi2019}, thereby providing the long-range information transmission services with high capacity to MUs \cite{LPZhu2019}. However, the mobility of UAVs presents challenges in achieving accurate beam tracking and alignment to MUs. While GPS can provide the locations of MUs and UAVs, its deployment is expensive, and location errors can occur in mobility scenarios. In addition, environmental factors such as wind can affect the locations of UAVs even when they follow pre-set trajectories. To address these challenges, angle-aware beam-tracking algorithms have been proposed to achieve beam tracking for UAV-to-MU links with high accuracy. 

The existing angle-aware predictive beam-tracking algorithms that consider angles require the historical trajectory of UAVs, but their accuracy in predicting locations may not be satisfactory for different motion patterns. Furthermore, the reconstruction time of a beam is often overlooked, while beam coverage is a crucial factor in constructing available beams \cite{LYang2019,HLSong2021}. Previous works have used beam coverage to create wide beams for error domains caused by angular predictions or regions that the pre-set codebook cannot cover. Transmission interruption occurs when the target MU moves outside the beam coverage region. Moreover, the fixed time interval for reconstructing a beam cannot keep up with the speed of MUs. Therefore, it is crucial to reconstruct a beam based on the beam coverage in an adaptive mode.

In order to ensure that the target MU remains within the coverage of the constructed beam, this study employs a combination of position-known assisted UAVs (A-UAV) and position-unknown assisted UAVs (U-UAV). This approach facilitates efficient information transmission to MUs with beam tracking. A novel bi-directional angle-aware beam-tracking with adaptive beam reconstruction (BAB-AR) algorithm is proposed to accurately reconstruct the beam based on the movement of the MU, thereby preventing beam misalignment due to high speed of both MUs and UAVs. Thus, the BAB-AR algorithm can mitigate communication interruptions that may occur due to reconstructing a beam with a fixed time interval. The main contributions of this paper are outlined below. 

\begin{itemize}
    \item A modified global dynamic crow search algorithm (GDCSA) is proposed for locating U-UAVs without GPS and for realizing accurate angle-aware beam tracking within the UAV-BSs. The GDCSA incorporates a predetermined optimal point set and a dynamical location updating mechanism to achieve fast convergence to the optimal solution. A-UAVs can cooperatively locate the U-UAVs using GDCSA, avoiding the need of high-dimensional computation regardless of any UAV's moving patterns.
    \item A Gaussian process regression (GPR) model is employed to predict the azimuth and elevation angles of the target MU based on the U-UAV's locations, which eliminates frequent pilot overhead. Furthermore, a time interval adjustment mechanism (TIAM) is proposed to calculate an adaptive time interval for reconstructing the beam according to the movement of MUs and the beam width.
    \item The performance of the proposed BAB-AR algorithm is verified under different movement patterns. Simulation results show that the angle relative error of the GDCSA is within 0.2 under different motion patterns. Furthermore, BAB-AR achieves higher SNR and transmit rate at MUs compared to that of the angle-aware beam tracking with machine model. When MU moves slowly, energy-efficiency at MU can be improved approximately 136.46\% and 215.99\% compared to that of the fixed time interval mode and the codebook-based algorithm, respectively.
\end{itemize}

The rest of this paper is organized as follows. Section II provides an overview of existing literatures on beam tracking for MUs. Section III presents the problem model and channel model used in the proposed scenarios. Section IV describes the process of the proposed algorithm in detail. Section V discusses the results of the simulation experiment, and Section VI concludes this study.

\section{Related work}
In this study, UAV-BSs are employed to enable accurate beam tracking for MUs. It is crucial to prompty reconstruct beams for MUs in order to prevent communication interruptions resulting from the high speed of both UAVs and MUs. The following prior works have explored various algorithms to update beam vectors and quickly reconstruct beams to accommodate the mobility of UAVs and MUs.  The characteristics of the existing algorithms and this study are summarized in Table \ref{table1}.

Some studies have focused on the codebook-based beam-tracking algorithms and developed various hierarchical multi-resolution codebooks to mitigate the impact on beam alignment caused by the high mobility of both MUs and UAVs \cite{LYang2019,WZZhang2020,JLZhang2020}. These algorithms initially designed and established the codebook, after which UAVs and MUs can select the optimal beam vector from the fixed codebook. The codebook-based beam-tracking algorithm minimizes the time required for beam training and beam vector selection. However, the fixed codebook lacks the ability to independently and promptly adapt to align with the target MU, resulting in suboptimal beam gain. Moreover, despite the continuity of the codebook, uncovered regions always exist due to the difference between the spatial frequency and geographic domains, which leads to communication interruptions between the UAVs and MUs \cite{SHHyun2022}.

Another solution is the beam vector-optimized beam-tracking algorithms. An alternating interference suppression algorithm \cite{LPZhu2020} was proposed to maintain UAVs' communication. By optimizing the beam vector for both the link between the BS and UAVs and the link between UAVs and MUs. This approach maximized the beam gain of the target signal while effectively mitigating interference. In \cite{XChen2020}, the least square algorithm was employed to optimize the beam vectors for UAVs and MUs, resulting in a cooperative sensing UAV network that maximized the beam gain of the target signal between UAVs and MUs. To enhance the accuracy of the optimal beam vector, an artificial bee colony algorithm was used to obtain a suitable beam vector in \cite{ZYXiao2020}. However, beam alignment from UAVs to MUs cannot be achieved quickly due to the impact of the number of antennas on the complexity and time required to solve the optimization problem in a 3D beam-tracking scenario. In addition, the number of iterations of the optimization algorithms can also affect the beam alignment. 

The beam vector is a crucial parameter for enhancing the accuracy of beam tracking in UAV-to-MU links. The time delay in finding the optimal beam vector significantly affects the accuracy of the beams. In addition, pilot overhead for channel estimation leads to energy consumption during both codebook-based and beam vector-optimized beam-tracking algorithm. To address these challenges, several studies have focused on angle of departure (AoD) and angle of arrival (AoA) in calculating beam vectors and proposed angle-aware predictive beam tracking based on localization. These algorithms enable the construction of a efficient beam-tracking mechanism without the need for pilot overhead between UAVs and MUs for channel estimation.

In \cite{CLiu2021}, a long short-term memory-based recurrent neural network (LRNet) was designed to predict the location of UAVs and calculate the corresponding angles. In \cite{YHuang2020}, the UAV motion was viewed as a linear motion, and a modified least squares search was used to obtain the spatial angle from UAV to BS, thereby achieving beam tracking in the BS-to-UAV links. In \cite{HLSong2021-R}, Kalman filtering was used to predict the spatial angle between the UAV and BS. This approach provided less training overhead and high beam alignment accuracy for beam tracking between BS and UAV. In \cite{JLZhang2019}, GPR was utilized to predict the locations of UAVs. However, the time delay between BS and UAVs impacts the accuracy of beam tracking from UAVs to MUs. To mitigate the effect of predictive angle errors on beam alignment, an algorithm for wide beam construction was proposed in \cite{HLSong2021}. However, this approach reduces the beam gain received at MUs. Various models such as SVM, Kalman filter, and Bayesian perspective were used to predict spatial angles of MUs by BS \cite{SJayaprakasam2017,MYang2021,WJYuan2021}. These algorithms always used historical trajectories to train the model. However, the trained model cannot accurately predict the UAV trajectory due to significant yet random impacts caused by the wind and other environmental factors. Nevertheless, the dynamic time interval for beam reconstruction has been overlooked, which results in an interruption in communication between the target MU and BS when the MU moves rapidly.
  
\begin{table*}
\caption{\sethlcolor{-blue}Comparisons of the existing methods and this work}
\label{table1}
\renewcommand\arraystretch{1.55}
\resizebox{1.0\linewidth}{!}{
\centering
\begin{tabular}{|c|c|c|c|c|c|c|} 
\hline
Methods                                                                                                       & Reference & UAV cooperation &  \begin{tabular}[c]{@{}c@{}}Spatial angle\\ prediction \end{tabular}& \begin{tabular}[c]{@{}c@{}}Dynamic beam \\ reconstruction\end{tabular} & \begin{tabular}[c]{@{}c@{}}Real-time \\communication\end{tabular} & Low complexity  \\ 
\hline
\multirow{3}{*}{\begin{tabular}[c]{@{}c@{}}Codebook based \\beam-tracking algorithms\end{tabular}}            & {[}13]    & \texttimes               & \texttimes                        & \checkmark                           & \texttimes                                                                 & \checkmark               \\ 
\cline{2-7}
                                                                                                             & {[}15]    & \texttimes               & \texttimes                        & \texttimes                           & \texttimes                                                                 & \checkmark               \\ 
\cline{2-7}
                                                                                                             & {[}17]    & \texttimes               & \texttimes                        & \texttimes                           & \texttimes                                                                 & \checkmark               \\ 
\hline
\multirow{3}{*}{\begin{tabular}[c]{@{}c@{}}Beam vector optimized \\with optimization algorithms\end{tabular}} & {[}18]    & \texttimes               & \texttimes                        & \texttimes                           & \texttimes                                                                 & \texttimes               \\ 
\cline{2-7}
                                                                                                             & {[}19]    & \checkmark               & \texttimes                        & \texttimes                           & \texttimes                                                                 & \texttimes               \\ 
\cline{2-7}
                                                                                                             & {[}20]    & \texttimes               & \texttimes                        & \texttimes                           & \texttimes                                                                 & \texttimes               \\ 
\hline
\multirow{3}{*}{Angle-aware beam-tracking algorithms}                                                                    & {[}21]    & \texttimes               & \checkmark                        & \texttimes                           & \checkmark                                                                 & \checkmark               \\ 
\cline{2-7}
                                                                                                             & {[}23]    & \texttimes               & \checkmark                        & \texttimes                           & \checkmark                                                                 & \checkmark               \\ 
\cline{2-7}
                                                                                                             & {[}14]    & \texttimes               & \checkmark                        & \texttimes                           & \checkmark                                                                 & \checkmark               \\ 
\hline
BAB-AR                                                                                                       & This work & \checkmark               & \checkmark                        & \checkmark                           & \checkmark                                                                 & \checkmark               \\
\hline
\end{tabular}
}
\end{table*}
\section{System model}
\subsection{Network model}
\begin{figure}[H]
    \centering
    \includegraphics[width=3in]{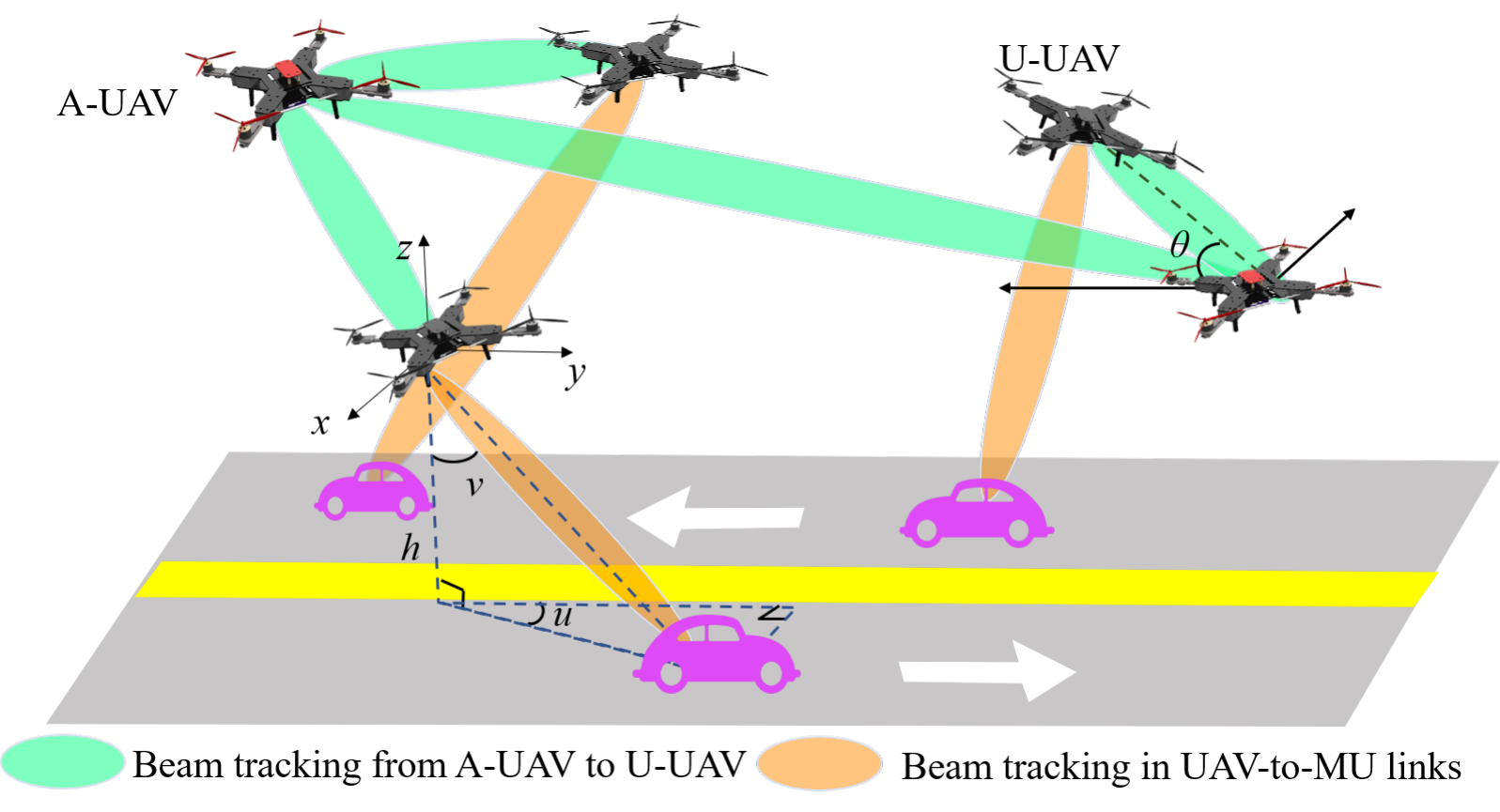}
    \caption{Communication model between UAVs and MUs.}
    \label{fig1}
\end{figure}

The beam-tracking system in this study is comprised of beam-tracking among UAV-BSs and UAVs to MUs. Each UAV can follow its own pre-set trajectory. For simplify, all UAVs are assumed to be on the same plane, and MUs are considered as cars that are always in motion. As illustrated in Fig.\ref{fig1}, the flight area of UAV-BSs is defined as $[(0, 0), (x_{max} , y_{max} )]$, where $k$ position-known assisted UAVs (A-UAV) and $m$ position-unknown assisted UAVs (U-UAV) at altitude $h$ as UAV-BSs are assumed to provide the service of beam tracking to MUs. The A-UAVs, which are randomly distributed in the area, are equipped with high-precision GPS and $N = N_{ULA}$ uniform line array (ULA). The U-UAVs are equipped with $N = N_{x} \times N_{y}$ uniform planar array (UPA) to provide beam-tracking for MUs \cite{LYang2019}. They also have a single antenna to receive information from A-UAVs.
\subsection{Channel model}
The mmWave channel used in this study is typically sparse in space, and the AoD and AoA change as UAVs and MUs move \cite{YJKim2018,WJYuan2021}. To achieve adaptive beam tracking among UAV-BSs, the position unknown U-UAV sends reference signals such as demodulation reference signals (DMRS) to locate themselves \cite{YHuang2020,MJiang2016}. The proposed scenario has two beamforming approaches: 2D beamforming among UAV-BSs and 3D beamforming towards MUs.
\subsubsection{2D beamforming from A-UAVs to U-UAVs}
As UAV-BSs, the locations of the A-UAV$_{k}$ and the U-UAV$_{i}$ are denoted as $(x_{k}, y_{k})$  and $(x_{i}, y_{i})$, respectively, under a 2D coordinate system. Then the spatial angle $\theta_{i, k}$ of the A-UAV$_{k}$ relative to the U-UAV$_{i}$ can be expressed as follows:
\begin{equation}
   {{\theta }_{i,k}}\text{=arctan}\frac{{{y}_{i}}-{{y}_{k}}}{{{x}_{i}}-{{x}_{k}}}
   \label{eq1}
\end{equation}
The direction vector ${{a}_{i,k}}\in {{\mathbb{C}}^{{{N}_{ULA}}\times 1}}$ of the reference signal received at the A-UAV$_{k}$ can be expressed as follows:
\begin{equation}
   {{a}_{i,k}}=\left[ 1,{{e}^{-j\cos {{\theta }_{i,k}}}},...,{{e}^{-(N_{ULA}-1)j\cos {{\theta }_{i,k}}}} \right]
   \label{eq2}
\end{equation}
In this study, the Rayleigh fading model, which is usually used for LoS paths, is used between the A-UAV$_{k}$ and the U-UAV$_{i}$ \cite{HLSong2021,CLiu2021}. Then the channel matrix H$_{i,k}$ between the A-UAV$_{k}$ and the U-UAV$_{i}$ can be expressed as follows:
\begin{equation}
   {{H}_{i,k}}=\frac{c \rho }{4\pi {{f}_{c}} {{d}_{i,k}}}
   \label{eq3}
\end{equation}
where $f_{c}$ is the operating frequency, $c$ is the speed of light, $\rho $ is the channel fading parameter, and $d_{i, k}$ represents the straight-line distance between the A-UAV$_{k}$ and the U-UAV$_{i}$. $d_{i, k}$ can be calculated as follows:
\begin{equation}
{{d}_{i,k}}=\sqrt{{{({{x}_{i}}-{{x}_{k}})}^{2}}+{{({{y}_{i}}-{{y}_{k}})}^{2}}}
   \label{eq4}
\end{equation}
The received signal at the A-UAV$_{k}$ equipped with a ULA can be expressed as follows:
\begin{equation}
\begin{split}
  {{r}_{i,k}}&= \rho  {{H}_{i,k}} {a}_{i,k}^{H} s_{i,k}+n \\ 
 &=\frac{ c \rho }{4\pi  {{f}_{c}} {{d}_{i,k}}} {a}_{i,k}^{H} s+n
\label{eq5}
\end{split}
\end{equation}
Let $s_{i,k}$ denote the reference signal sent from the U-UAV$_{i}$ to the A-UAV$_{k}$, and ${{n}_{0}}\sim CN\left( 0,{{\sigma }^{2}} \right)$ is Gaussian white noise, where $\sigma^{2}$ set as 1. The signal received at U-UAV$_{i}$ from A-UAV$_{k}$ can be expressed as follows:
\begin{equation}
\begin{split}
  {{r}_{k,i}}&=\sqrt{N_{ULA}} \rho {w_{i,k}} {{H}_{i,k}} {a}_{k,i}^{H} s_{i,k}+n \\ 
 &=\frac{ \sqrt{N_{ULA}} c \rho }{4\pi  {{f}_{c}} {{d}_{i,k}}}{w_{i,k}} {a}_{k,i}^{H} s+n
\label{eq6}
\end{split}
\end{equation}
Given that the UAVs are in the same plane, the spatial angle from U-UAV$_i$ to A-UAV$_k$ is equal to the spatial angle from A-UAV$_k$ to U-UAV$_i$, and the direction vector $a_{i,k}$ is equal to $a_{i,k}$ as shown in Eq. (\ref{eq2}). In addition, $s_{k,i}$ in Eq. (\ref{eq6}) denotes the signal sent from the A-UAV$_k$ to the U-UAV$_i$, and $w_{i,k}$ is the beam vector from the A-UAV$_k$ to the U-UAV$_i$, which can be expressed as follows:
\begin{equation}
\label{eq7}
   \begin{split}
   {{w}_{i,k}}&=\frac{1}{\sqrt{{{N}_{ULA}}}} 1,{{e}^{-j\cos {{{\hat{\theta }}}_{i,k}}}},...,{{e}^{-(M-1)j\cos {{{\hat{\theta }}}_{i,k}}}}
 \end{split}
\end{equation}
where ${{\hat{\theta }}_{i,k}}$ is the optimized angle from U-UAV$_i$ to A-UAV$_k$ which can be calculated as follows:
\begin{equation}{{\hat{\theta}}_{i,k}}\text{=arctan}\frac{{\hat{y}_{i}}-{{y}_{k}}}{{\hat{x}_{i}}-{{x}_{k}}}
\label{eq8}
\end{equation}
The calculation of the optimized location of U-UAV$_{i}$ $(\hat{x}_{i},\hat{y}_{i})$ is introduced in Section IV. 
\subsubsection{3D beamforming from U-UAV to the target MU}
The location coordinate of the target MU$_{l}$ is $(x_{l}, y_{l}, 0)$ and the U-UAV$_{i}$ is $(x_{i}, y_{i}, h)$ under a 3D coordinate system as shown in Fig. \ref{fig1}. Thus, the distance between the MU$_{l}$ and the U-UAV$_{i}$ can be expressed as follows:
\begin{equation}
    {{d}_{l}}=\sqrt{{{({{x}_{l}}-{{x}_{i}})}^{2}}+{{({{y}_{l}}-{{y}_{i}})}^{2}}+{{h}^{2}}}
    \label{eq9}
\end{equation}
The azimuth and elevation angles $u_l$ and $v_l$ from the MU$_l$ to the U-UAV$_i$ can be calculated by the locations of MU$_l$ as follows:
\begin{equation}
    {{u}_{l}}=\arctan \frac{{{y}_{l}}-{{y}_{i}}}{{{x}_{l}}-{{x}_{i}}}
    \label{eq10}
\end{equation}
\begin{equation}
    {{v}_{l}}=\arctan \frac{\sqrt{{{({{y}_{i}}-{{y}_{l}})}^{2}}+{{({{x}_{i}}-{{x}_{l}})}^{2}}}}{h}
    \label{eq11}
\end{equation}
After the A-UAV$_{k}$ predicts the spatial angle of the MU$_{l}$ and transmits data containing the locations of U-UAVs and MU$_l$ to the U-UAV$_{i}$, the U-UAV$_{i}$ equipped with the $N_{x} \times N_{y}$ UPA can provide beam-tracking to the MU$_{l}$. Additionally, the channel matrix $H_{l}$ between the U-UAV$_{i}$ and the MU$_{l}$ can be expressed as follows:
\begin{equation}
    {{H}_{l}}=\frac{c {{\rho }_{l}}}{4\pi  {{f}_{c}} {{d}_{l}}}
    \label{eq12}
\end{equation}
where $\rho _{l}$ is the channel fading between the U-UAV$_{i}$, the MU$_{l}$, $d_{l}$ denotes the distance between the U-UAV$_{i}$ and the MU$_{l}$, and the signal direction vector ${{a}_{l}}\left( {{u}_{l}},{{v}_{l}} \right)$ emitted from the U-UAV$_{i}$ to the MU$_{l}$ can be denoted as follows:
\begin{equation}
 \label{eq13}
 \begin{split}
    {{a}_{l}}({{u}_{l}},{{v}_{l}})=( 1,{{e}^{j2\pi c/{{f}_{c}}( \cos {{v}_{l}}+\sin {{v}_{l}} )\sin {{u}_{l}}}},..., \\ {{e}^{j2\pi c/{{f}_{c}}( ( {{N}_{x}}-1 )\cos {{v}_{l}}+( {{N}_{y}}-1 )\sin {{v}_{l}} )\sin {{u}_{l}}}} )
\end{split}
\end{equation}
Thus the signal received by the MU$_{l}$ can be expressed as follows:
\begin{equation}
    \begin{split}  
   {{r}_{i,l}}&=\sqrt{{{N}_{x}}\times {{N}_{y}}} {{w}_{l}}({{{\hat{u}}}_{l}},{{{\hat{v}}}_{l}}) {{H}_{l}} {{a}_{l}}{{({{u}_{l}},{{v}_{l}})}^{H}}+n \\ 
 &={{w}_{l}}({{{\hat{u}}}_{l}},{{{\hat{v}}}_{l}}) \frac{\sqrt{{{N}_{x}}\times {{N}_{y}}} c {{\rho }_{l}}}{4\pi  {{f}_{c}} {{d}_{l}}} {{a}_{l}}{{({{u}_{l}},{{v}_{l}})}^{H}}+n 
\label{eq14}
\end{split}
\end{equation}
where $w_{l}$ is the beam vector from the U-UAV$_{i}$ to the MU$_{l}$, which can be denoted as follows:
\begin{equation}
\begin{split}
    {{w}_{l}}({{\hat{u}}_{l}},{{\hat{v}}_{l}})=\frac{1}{\sqrt{{{N}_{x}}\times {{N}_{y}}}}( 1,{{e}^{j2\pi c/{{f}_{c}}( \cos {{{\hat{v}}}_{l}}+\sin {{{\hat{v}}}_{l}} )\sin {{{\hat{u}}}_{l}}}},\\...,{{e}^{j2\pi c/{{f}_{c}}( ( {{N}_{x}}-1 )\cos {{{\hat{v}}}_{l}}+( {{N}_{y}}-1)\sin {{{\hat{v}}}_{l}} )\sin {{{\hat{u}}}_{l}}}})
\end{split}
\label{eq15}
\end{equation}
where ${{\hat{u}}_{l}}$ and ${{\hat{v}}_{l}}$ are the predicted azimuth and elevation angles from the MU$_{l}$ to the U-UAV$_{i}$. Moreover, the half power beam width (HPBW) of the beam in azimuth plane and elevation plane can be calculated as follows:
\begin{equation}
    {{\theta }_{u}}\left( u,v \right)=\frac{1}{\cos u\sqrt{\Delta \theta _{u}^{-2}{{\cos }^{2}}v+\Delta \theta _{v}^{-2}{{\sin }^{2}}v}}
    \label{eq16}
\end{equation}
\begin{equation}
    {{\theta }_{v}}\left( u,v \right)=\frac{1}{\sqrt{\Delta \theta _{u}^{-2}{{\sin }^{2}}v+\Delta \theta _{v}^{-2}{{\cos }^{2}}v}}
    \label{eq17}
\end{equation}
where $\Delta {{\theta }_{u}}$ and $\Delta {{\theta }_{v}}$ are the HPBW of the beam constructed by linear array with $N_{x}$ and $N_{y}$ antennas, respectively, the calculation can be expressed as follows:
\begin{equation}
    \Delta {{\theta }_{u}}=\frac{\lambda \times 2}{{{N}_{x}}}
    \label{eq18}
\end{equation}
\begin{equation}
    \Delta {{\theta }_{v}}=\frac{\lambda \times 2}{{{N}_{y}}}
    \label{eq19}
\end{equation}
\section{Bi-directional angle-aware beam-tracking with adaptive beam reconstruction algorithm}
Existing angle-aware beam-tracking algorithms have been developed for scenarios involving the communication between BS and MUs or BS and UAVs \cite{HLSong2021} \cite{CLiu2021,YHuang2020,HLSong2021-R,JLZhang2019,MYang2021,WJYuan2021,MJiang2016}. However, these algorithms fail to address the fixed height of the BS, which results in a lack of LoS links. The time delay in transmitting signals from the BS to UAVs also affects the accuracy of UAV localization, and existing algorithms struggle to maintain stability across different movement patterns of UAVs. Consequently, the accuracy of beam tracking from UAVs to MUs remains unsatisfactory. In addition, the time interval for predicting the angle used during beam reconstruction for high-speed MUs is not considered. To address these limitations, this study proposes a bi-directional angle-aware beam-tracking with adaptive beam reconstruction (BAB-AR) algorithm. 

\begin{figure}
    \centering
    \includegraphics[width=3in]{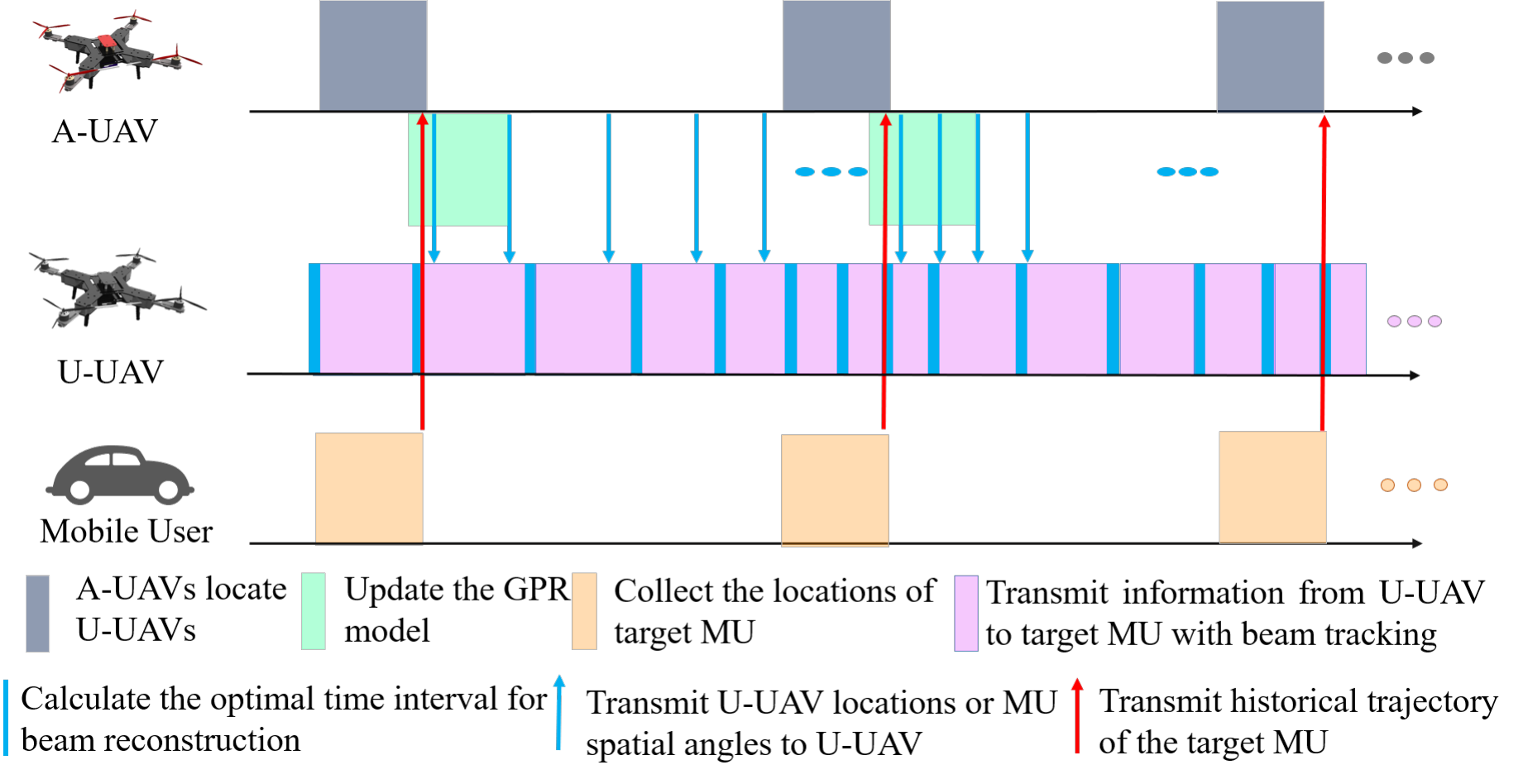}
    \caption{Time frame of the BAB-AR algorithm.}
    \label{fig2}
\end{figure}

The communication cycle in BAB-AR is divided into two parts: angle-aware beam-tracking within UAV-BSs and beam tracking from UAVs to MUs. As Fig. \ref{fig2} shows, the two processes of A-UAVs, localization for U-UAVs and collection for the historical trajectory of the target MU, run in parallel. During U-UAV localization phase, GDCSA is proposed to accurately locate U-UAVs without GPS for fast convergence. Once the localization of U-UAVs is finished, the historical spatial angles from MU to U-UAV can be obtained using Eq. (\ref{eq10}) and Eq. (\ref{eq11}), and the GPR model can be trained by A-UAV for the target MU. As the model and predicted locations of U-UAV are updated, beam tracking within the UAV-BSs can be carried out as described in Section III. B. Furthermore, the real-time information, such as the optimized U-UAVs locations and the predicted angles from the target MU to U-UAV, can be transmitted from A-UAVs to U-UAVs. \par
\begin{figure}
    \centering
    \includegraphics[width=3in]{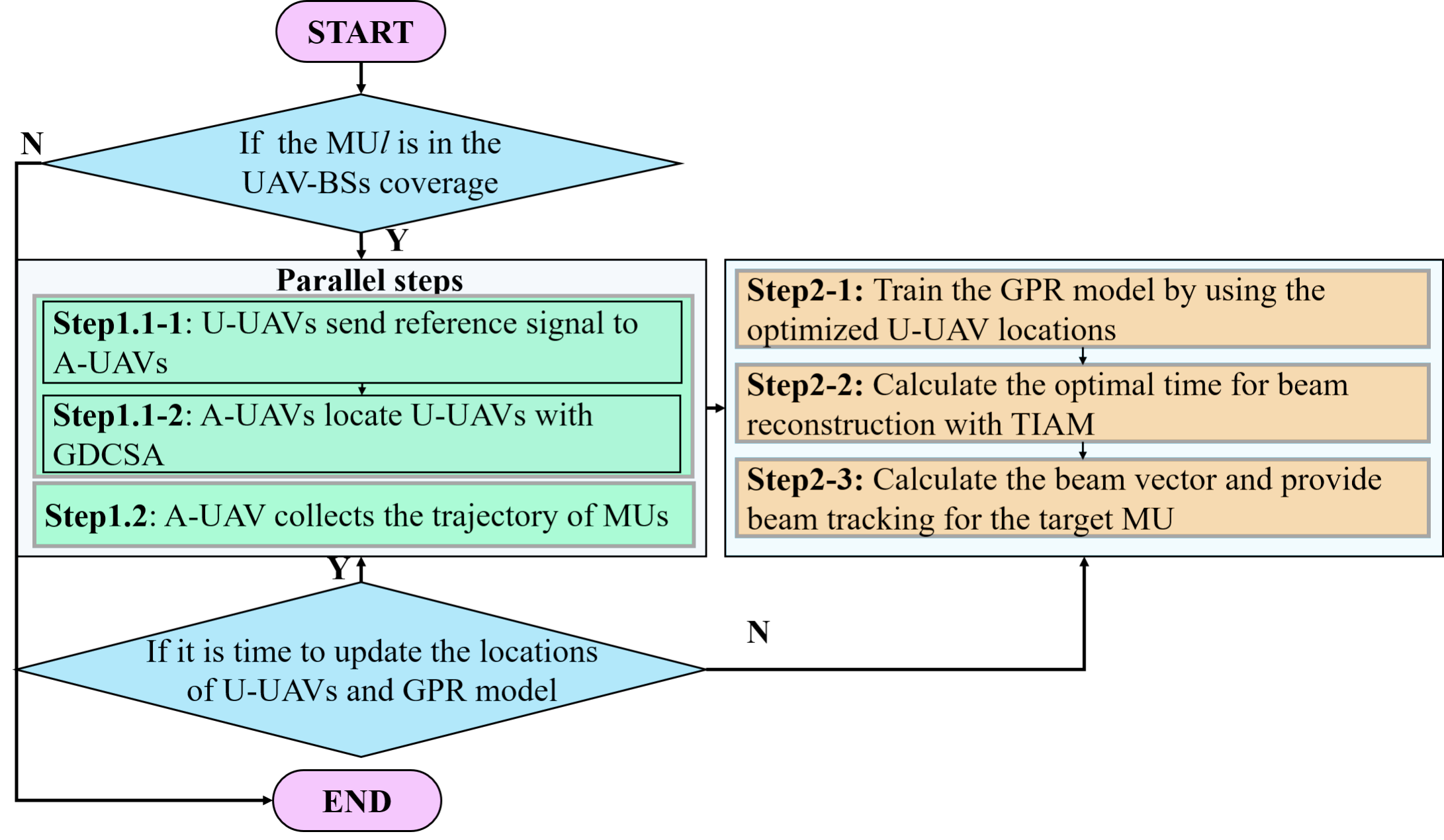}
    \caption{Flow chart of BAB-AR algorithm.}
    \label{fig3}
\end{figure}

 During the communication between UAV and MU, the trained GPR model is used to predict the spatial angles from the MU$_l$ to the U-UAV$_i$ at time $t_{n-1}+\Delta t_{n}^{*}$, where $\Delta t_{n}^{*}$ is the dynamic time interval calculated by TIAM for beam reconstruction. With the predicted locations of U-UAV$_i$ and spatial angle of MU$_l$, the U-UAV$_i$ can provide timely beam tracking for the high-speed MU$_l$. In this phase, only location information needs to be collected, thus, the pilot overhead caused by channel estimation can be decreased \cite{CLiu2021}. In addition, A-UAV updates the locations of U-UAVs every $t_{check}$, and the efficiency of the model is checked with the new trajectory of the target MU. The flow chart of BAB-AR is shown in Fig. \ref{fig3}.
\subsection{Localization of UAV-BSs via the proposed GDCSA}
The accurate localization of UAV-BSs using GPS for beam construction is costly. Moreover, existing angle-aware beam-tracking algorithms for the UAV-BSs usually require the historical trajectory of UAVs to perform location prediction. However, the locations of UAVs are always affected by environmental factors such as wind, and the historical trajectory information may not be accurate when training a model to predict UAV location. To address this issue, the GDCSA is designed with low computing complexity to cooperatively locate U-UAVs by A-UAVs without the historical trajectory data. A-UAVs can calculate the position of U-UAV collaboratively by utilizing the signal transmitted from the U-UAV to the A-UAVs. Therefore, the accuracy of the predictive location cannot be affected by the moving patterns or the shaking of UAV, since the localization process is independent of the U-UAV historical trajectory.
\subsubsection{GDCSA}
Crow search algorithm (CSA) is a swarm intelligence optimization algorithm with fast convergence \cite{AAskarzadeh2016}. In CSA, each crow's position in the search space represents a feasible solution, and all crows gradually explore the search space by leveraging the objective function's value as a reference. The iterative process eventually leads to a global optimal solution.

In CSA, crow $i$ randomly selects a crow $j$ from the population as the learning target at each iteration, and then the location of crow $i$ is updated as follows:
\begin{equation}
    X_{i}^{t+1}=\left\{ \begin{aligned}
  & X_{i}^{t}+{{r}_{i}}\times fl_{i}^{t}\times \left| m_{i}^{t}-X_{i}^{t}  \right|, &{{r}_{i}}\ge AP_{i}^{t} \\ 
 & \text{a random position},  &\text{otherwise}
\end{aligned} 
\right.
\label{eq20}
\end{equation}
where, $AP_{i}^{t}$ denotes the probability that the crow $i$ found the optimal crow at the $t$ iteration. $fl_{i}^{t}$ denotes the flight distance of the crow$_i$  at the $t$ iteration, and $r_{i}$ is a random number between 0 and 1. One limitation of the original CSA is that the initial positions of crows are generated randomly, which can lead to an uneven distribution of solutions in the search space. Moreover, in the original CSA, the value of $AP_{i}^{t}$ remains constant throughout the algorithm, which can slow down the convergence speed.

A GDCSA is proposed to increase the global search ability of crows and improve the optimization accuracy. In this approach, the optimal point set is used to initialize the crows, which can be calculated using Eq. (\ref{eq21}). The crows then disperse evenly across the search space, as depicted by the stars in Fig. \ref{fig4}. Thus, the global search ability of the cows is enhanced, which leads to superior optimization results\cite{HYan2016}. 
\begin{equation}
\begin{split}
    {{P}_{n}}( k )=\{ (\{ r_{1}^{( n )}\times k \},\{ r_{2}^{( n )}\times k \},\cdot \cdot \cdot ,\\\{ r_{s}^{( n)}\times k \} ),1\le k\le n \}
\end{split}
\label{eq21}
\end{equation}
where $r_{i}$ is set as follows:
\begin{equation}
    {{r}_{i}}=\left\{ 2\cos \frac{2\pi i}{p},1\le i\le s \right\}
    \label{eq22}
\end{equation}
where $s$ is the number of points, and $p$ is the smallest prime number which satisfies the condition as follows:
\begin{equation}
    \frac{p-3}{2}\ge s
    \label{eq23}
\end{equation}
A dynamic parameter $AP_{i}^{t}$ adapted to the number of iterations is likewise designed to enhance the rate of convergence.
\begin{equation}
    AP_{i}^{t}=\beta  {{\left( \frac{iter\_\max -t}{iter\_\max } \right)}^{2}}
    \label{eq24}
\end{equation}
where $iter\_\max$ is the initial maximum iteration number of the algorithm, and $t$ is the current number of iterations. According to Eq. (\ref{eq24}), the initial value of $AP$ is substantial, but it gradually decreases as the iterations continue. 
\subsubsection{U-UAVs localization based on GDCSA and beam construction}
\begin{figure}
    \centering
    \includegraphics[width=3in,height=1.5in]{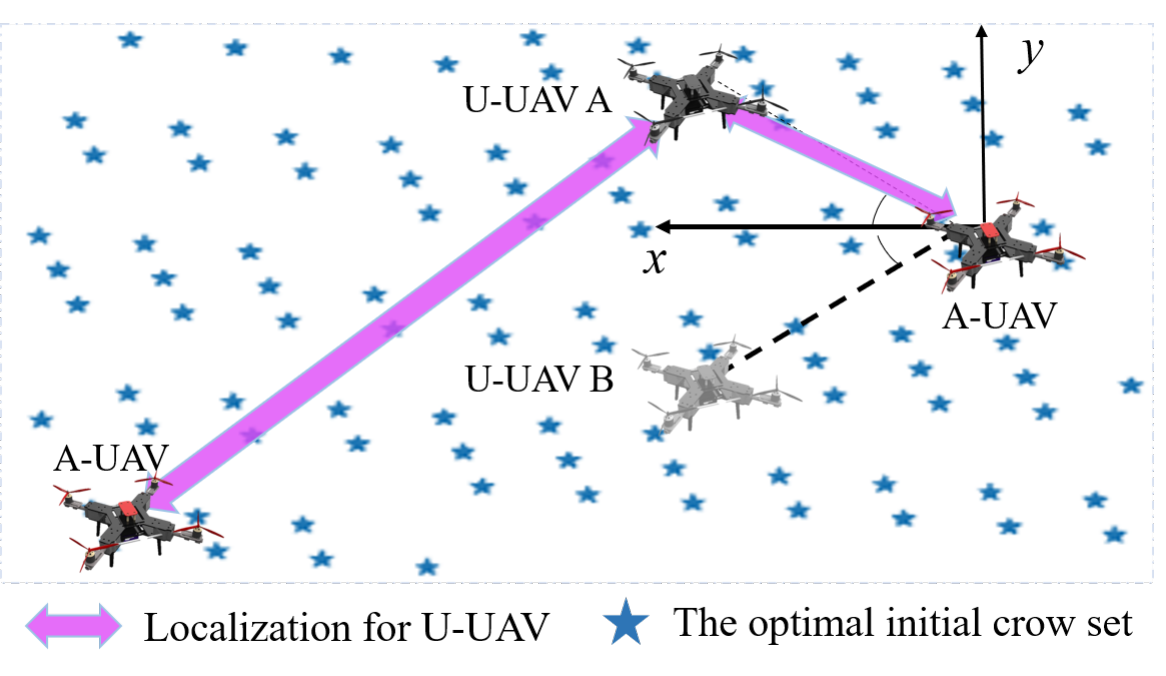}
    \caption{$k$ A-UAVs jointly locating U-UAVs.}
    \label{fig4}
\end{figure}

In this section, the GDCSA is used to locate U-UAVs by A-UAVs. The scenario involves $k$ A-UAVs and $m$ U-UAVs functioning as UAV-BSs. Based on the received signal at A-UAVs from U-UAVs, A-UAVs locate U-UAVs with GDCSA cooperatively. During the localization phase, A-UAVs with known locations communicate with each other through beam tracking. However, if U-UAVs \textit{A} and \textit{B} are symmetrical about the \textit{x}-axis, as illustrated in Fig. \ref{fig4}, then the GDCSA cannot accurately optimize the location of U-UAV when only one A-UAV is involved. Because the received signal strength at A-UAV from U-UAVs \textit{A} and \textit{B} is identical, which results in two possible solutions that satisfy the objective function. Therefore, it is crucial to have more than two A-UAVs for accurate localization.

In this study, $k$ A-UAVs are employed to cooperatively locate the U-UAV$_{i}$ in the target area. The main steps of constructing beam using GDCSA within UAV-BSs can be outlined as follows.
\begin{itemize}
    \item \textbf{Step 1:} The U-UAV$_i$ sends the reference signal to the $k$th A-UAV. Then the information containing the strength of the reference signal received from U-UAV$_i$ can be transmitted between A-UAVs with beam tracking.
\end{itemize}
\begin{itemize}
    \item \textbf{Step 2:} The optimal set is set as the initial population within the flight area of UAV-BSs $[(0, 0), (x_{max}, y_{max})]$.
\end{itemize}
\begin{itemize}
    \item \textbf{Step 3:} A-UAVs can cooperatively optimize the location of U-UAV$_i$ based on the objective function, which is defined as follows: 
  \begin{equation}
      \resizebox{0.39\textwidth}{!}{$y=\min \frac{1}{2}\sum\limits_{k=0}^{K-1}{\left| {{r}_{i,k}}-\frac{c}{4\pi {{f}_{c}}\sqrt{{{\left( \hat{x}-{{x}_{k}} \right)}^{2}}+{{\left( \hat{y}-{{y}_{k}} \right)}^{2}}}} {{a}_{i,k}} \right|}$}
    \label{eq25}
\end{equation}
    where $r_{i,k}$ is the strength of the actual signal received by the A-UAV$_k$, and $\frac{c}{4\pi {{f}_{c}}\sqrt{{{\left( \hat{x}-{{x}_{k}} \right)}^{2}}+{{\left( \hat{y}-{{y}_{k}} \right)}^{2}}}} {{a}_{i,k}} $ is the signal strength calculated based on position information. Moreover, $a_{i,k}$ is the direction vector, which can be calculated using Eq. (\ref{eq2}) with the location of U-UAV$_i$ and A-UAV$_k$. 
\end{itemize}
\begin{itemize}
    \item \textbf{Step 4:} The A-UAV closest to the U-UAV$_i$ transmits the information to the U-UAV$_i$ by beam tracking within the UAV-BSs. The information mainly contains the U-UAV$_i$'s location and the predicted angles of the target MU.
\end{itemize}
\textbf{Remark 1}: During the optimization process, each crow seeks the optimal solution that minimizes the value of Eq. (\ref{eq25}). The optimum result $({{\hat{x}}_{k}},{{\hat{y}}_{k}})$ is considered as the location of the U-UAV$_i$ at the end of the iteration. Thus, the received signal will change with the shaking of U-UAV, and the location of U-UAV is only related to the signal strength received by the A-UAV, which means that the accuracy of U-UAV location cannot be affected by the UAV shaking. 

The pseudo-code of the angle-aware beam-tracking algorithm for UAV-BSs is listed in detail as below. In addition, A-UAVs updates the location of U-UAV every $t_{check}$.
\begin{algorithm}
\caption{Angle-aware beam-tracking with the GDCSA}
\label{alg:alg1}
\renewcommand{\algorithmicrequire}{\textbf{Input:}}
\renewcommand{\algorithmicensure}{\textbf{Output:}}
\begin{algorithmic}
        \REQUIRE The reference signal send by U-UAV$_{i}$, locations of A-UAVs, maximum number of iterations $iter\_max$, $fl$ and the number of population $NP$  
        \ENSURE The optimal location of U-UAV$_{i}$ $(\hat{x},\hat{y})$   
        
        \STATE  initialize the crows as the optimal point set within the flight area of UAV-BSs
        
        \WHILE{$t<iter\_max$}
        \FOR{each $i \in [1,NP]$}
            \STATE Randomly choose one of the crows as crow$_j$
            \STATE Update the value of $AP$ according to Eq. (\ref{eq24})
            \STATE Update the location of crow$_{iter}$ as Eq. (\ref{eq20})
            
        \ENDFOR
        \STATE Update the fitness of crows according to Eq. (\ref{eq25})
        \ENDWHILE
        \RETURN The optimal location of U-UAV$_{i}$ $(\hat{x},\hat{y})$  
    \end{algorithmic}
\end{algorithm}
\subsection{A beam-tracking method with GPR for MUs}
GPR algorithm is used in this study to predict the trajectory of the target MU for decreasing the communication frequency between A-UAV and MU and minimizing the pilot overhead caused by channel estimation. Given the temporal correlation of MUs’ trajectory, GPR algorithm is suitable for timely predicting the spatial angles of MUs. Consequently, an angle-aware beam-tracking algorithm is proposed to reconstruct a beam with an adaptive time interval for tracking MUs. A-UAVs continuously gather the locations of the target MU$_{l}$ and check the error between the real-time location-derived angle and the GPR-predicted angle every $t_{check}$. If the error surpasses ${{\sigma }_{check}}$, the A-UAV will update the GPR model using the collected data, as illustrated in Fig. \ref{fig2} and Fig. \ref{fig3}. 

The azimuth angle and the elevation angle from the MU$_{l}$ to the U-UAV$_{i}$ at time $t_{n}$ can be mathematically denoted as $(u[t_{n}], v[t_{n}])$, which can be modeled as follows:
\begin{equation}
  u[{{t}_{n}}]={{g}_{u}}({{t}_{n}})+{{\varepsilon }_{u}}[{{t}_{n}}]
\label{eq26}
\end{equation}
\begin{equation}
    v[{{t}_{n}}]={{g}_{v}}({{t}_{n}})+{{\varepsilon }_{v}}[{{t}_{n}}]
\label{eq27}
\end{equation}
where $t_{n}$ is the time for beam reconstruction. ${{\varepsilon }_{u}}[{{t}_{n}}]$ and ${{\varepsilon }_{v}}[{{t}_{n}}]$ are noise with variance $\sigma _{u}^{2}$ and $\sigma _{v}^{2}$, respectively. ${{g}_{u}}(\centerdot )$ and ${{g}_{v}}(\centerdot )$ are the unknown function models fitted by GPR. In this study, the spatial angles between the MU$_{l}$ and the U-UAV$_{i}$ during $T=(t_{1}, t_{2}, … , t_{n})$ is mapping to $[0, 1]$ by Eq. (\ref{eq28}). The converted dataset is used for training the unknown function models ${{g}_{u}}(\centerdot )$ and ${{g}_{v}}(\centerdot )$ fitted by GPR.  The calculation of $g_{u}(t_{n+1})$ for predicting the azimuth angle at time $t_{n+1}$ is introduced as follows:
\begin{equation}
    {{g}_{u}}\left( {{t}_{i}} \right)=\left( u\left( {{t}_{i}} \right)-\min \left( U \right) \right)/\left( \max \left( U \right)-\min \left( U \right) \right)
    \label{eq28}
\end{equation}
where the range of $i$ is $[1,n]$, $U$ is the real spatial angle, and $u(t_{i})$ is the spatial angle at $t_{i}$. In GPR, ${{g}_{u}}(\centerdot )$ is a random variable that satisfies a Gaussian distribution ${{g}_{u}}\left( t \right)\tilde{\ }GP\left( m\left( T \right),K\left( T,{{T}^{'}} \right) \right)$, which can be defined as follows:
\begin{equation}
    E \left( {{g}_{u}}\left( T \right) \right)=m\left( T \right)
    \label{eq29}
\end{equation}
\begin{equation}
    \operatorname{cov}\left( {{g}_{u}}\left( T \right) \right)=K\left( T,{{T}^{'}} \right)
    \label{eq30}
\end{equation}
where Radial Basis Function (RBF) is chosen as the kernel function to adapt the changes of the spatial angles $u$ of the MU$_l$, which can be defined as follows:
\begin{equation}
    K\left( T,{{T}^{'}} \right)=\exp \left( -\frac{{{\left\| T-{{T}^{'}} \right\|}^{2}}}{2{{\sigma }^{2}}} \right)
    \label{eq31}
\end{equation}
 \sethlcolor{-blue}The spatial angle $u[t_{n+1}]$ at $t_{n+1}$ can be predicted when the value of $g_{u}(t)$ is $t_{n+1}$. Although the distribution of $g_{u}(t_{n+1})$ is unknown, the random variable consisting of $g_{u}(t_{n+1})$ and $Y= g_{u}(T)$ satisfies a Gaussian distribution as demonstrated in follows:
\begin{equation}
\resizebox{0.43\textwidth}{!}{$\left( \begin{aligned}
  & Y \\ 
 & {{g}_{u}}\left( {{t}_{n+1}} \right) \\ 
\end{aligned} \right)\text{=}N\left( \left( \begin{aligned}
  & \mu \left( T \right) \\ 
 & \mu \left( {{t}_{n+1}} \right) \\ 
\end{aligned} \right),\left( \begin{aligned}
  & K\left( T,T \right)+\sigma _{u}^{2}I \\ 
 & K\left( {{t}_{n+1}},T \right) \\ 
\end{aligned} \right.\left. \begin{aligned}
  & \text{  }K\left( T,{{t}_{n+1}} \right) \\ 
 & \text{  }K\left( {{t}_{n+1}},{{t}_{n+1}} \right) \\ 
\end{aligned} \right) \right)$}
\label{eq32}
\end{equation}
The GPR model can calculate the posterior distribution of $g_{u}(t_{n+1})$, which can be expressed as follows:
\begin{equation}
    P\left( f\left( {{T}^{*}} \right)|Y,X,{{X}^{*}} \right)\text{=}N\left( {{\mu }^{*}},{{\sum }^{*}} \right)
    \label{eq33}
\end{equation}
where ${{\mu }^{*}}$ and ${{\sum }^{*}}$ can be expressed as follows: 
\begin{equation}
\resizebox{0.43\textwidth}{!}{${{\mu }^{*}}\text{=}K\left( {{t}_{n+1}},T \right) {{\left( K\left( T,T \right)+\sigma _{u}^{2}I \right)}^{-1}}\left( Y-\mu \left( T \right) \right)+\mu \left( {{t}_{n+1}} \right)$}
\label{eq34}
\end{equation}
 \begin{equation}
         \resizebox{0.43\textwidth}{!}{${{\sum }^{*}}\text{=}K\left( {{t}_{n+1}},{{t}_{n+1}} \right)-K\left( {{t}_{n+1}},T \right){{\left( K\left( T,T \right)+\sigma _{u}^{2}I \right)}^{-1}}K\left( T,{{t}_{n+1}} \right)$}
    \label{eq35}
\end{equation}
where $\sigma _{u}^{2}$ represents the noise variance of the training data. Bringing the predictions ${{\hat{u}}_{l}}, {{\hat{v}}_{l}}$ for the MU$_l$ into Eq. (\ref{eq15}), the beam vector for the link between the U-UAV$_{i}$ and the MU$_{l}$ can be calculated as follows:
\begin{equation}
\begin{split}
    w({{u}_{l}},{{v}_{l}})=(1,{{e}^{j2\pi c/{{f}_{c}}( \cos {{{\hat{v}}}_{l}}+\sin {{{\hat{v}}}_{l}})\sin {{{\hat{u}}}_{l}}}},...,\\{{e}^{j2\pi c/{{f}_{c}}(( {{N}_{x}}-1)\cos {{{\hat{\varphi }}}_{l}}+( {{N}_{y}}-1 )\sin {{{\hat{v}}}_{l}})\sin {{{\hat{u}}}_{l}}}})
\end{split}
    \label{eq36}
\end{equation}
Then the beam can be constructed from the U-UAV$_{i}$ to the MU$_{l}$. Meanwhile, the communication link can be established between UAV-BSs and MUs at time $t_{n+1}$.
\subsection{Adaptive beam reconstruction with TIAM}
\begin{figure}[H]
    \centering
    \includegraphics[width=2.5in]{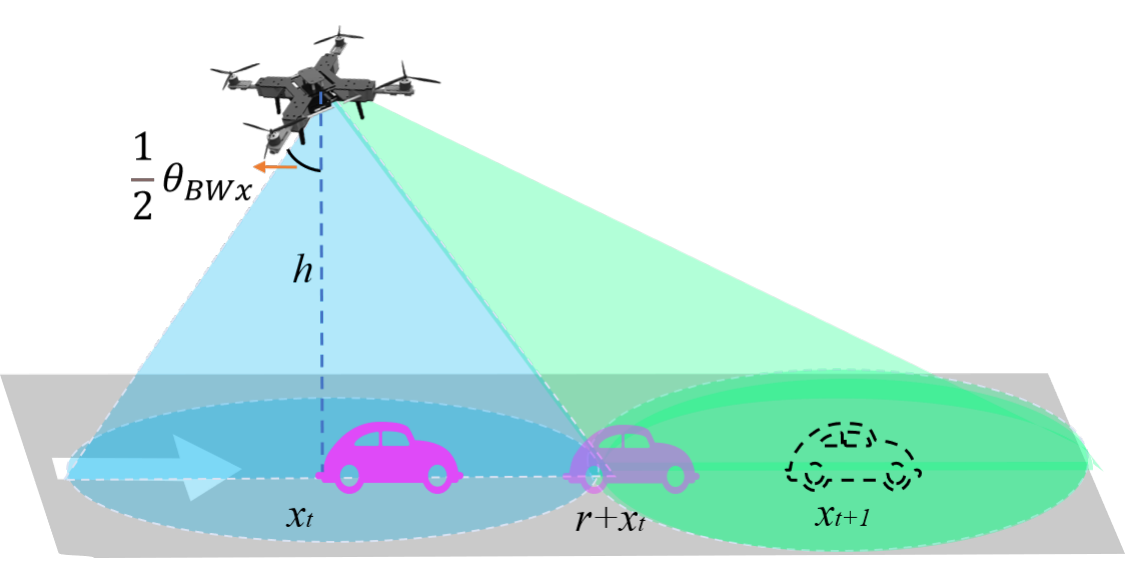}
    \caption{Beam reconstruction with fixed time interval.}
    \label{fig5}
\end{figure}

The beam can be constructed by the U-UAV$_{i}$ for the MU$_{l}$ once the angle is predicted by GPR. However, if the time interval $\Delta t$ for beam reconstruction cannot adapt the movement of the MU$_l$, the MUs may move out of the limited HPBW of the constructed beam. As shown in Fig. \ref{fig5}, when the target MU moves out, the beam should be rebuilt to cover the moving MU. In addition, a shorter time interval $\Delta t$ will result in frequent beam reconstruction, which leads to extra energy dissipation. Thus, a TIAM is proposed for the adaptive beam reconstruction during beam tracking from the U-UAV$_i$ to the MU$_l$, and the time interval $\Delta t$ can be set dynamically according to the moving state of the MU$_l$. Then the beam will be reconstructed at $t_n + \Delta t_{n}$, where $t_n$ represents the time of the last beam reconstruction. In addition, the moving state is determined by the angles predicted by GPR.
\begin{figure}[H]
    \centering
    \includegraphics[width=2.7in]{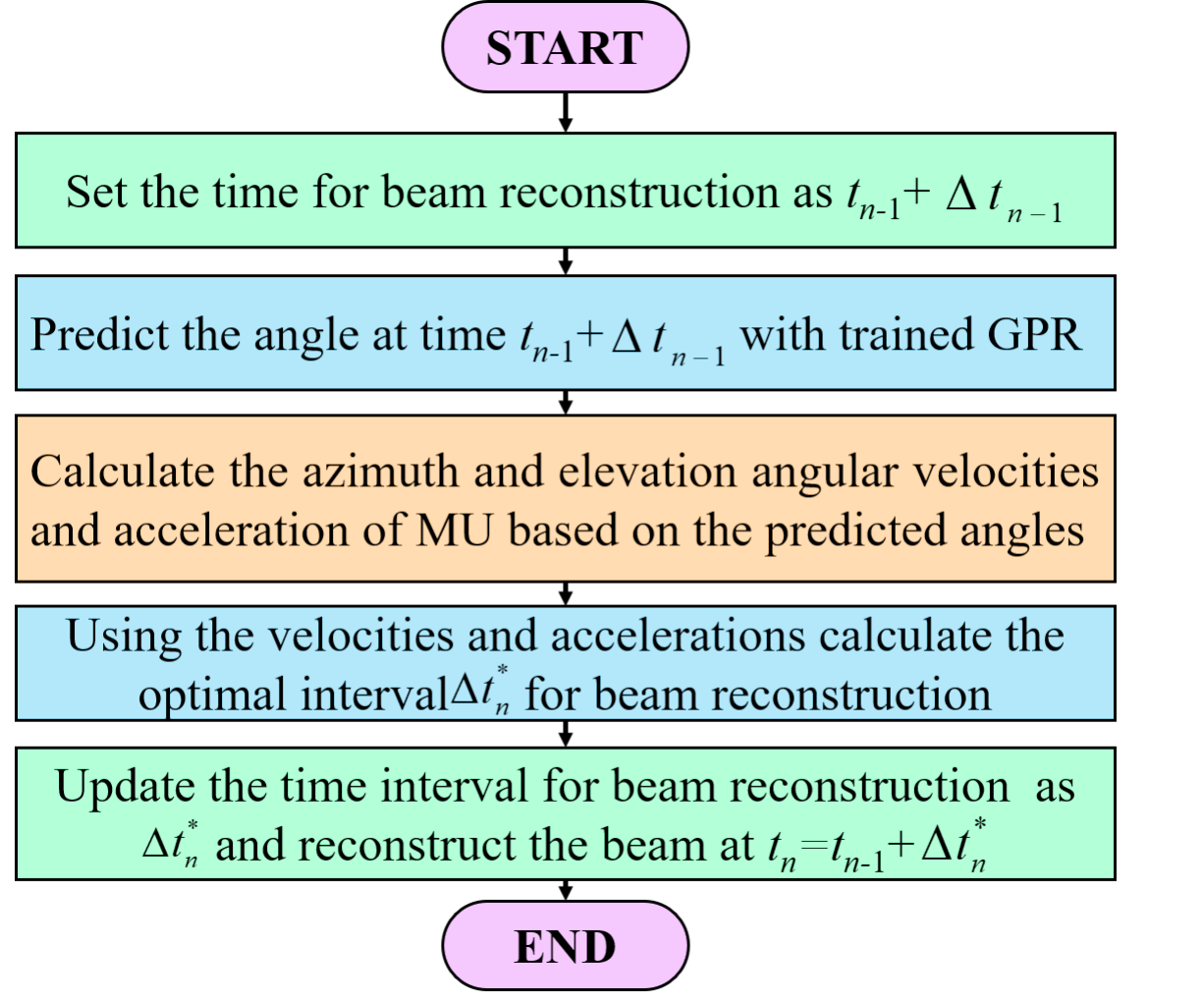}
    \caption{Flow chart of the proposed TIAM.}
    \label{fig6}
\end{figure}
As Fig. \ref{fig6} shown, the time interval $\Delta t_{n}$ for beam reconstruction is set as $\Delta t_{n-1}$ at the beginning of TIAM, when $n=0$, $\Delta t_{n}$ is set as the time interval of the training data (MU$_l$’s trajectory). The movement of MU$_l$ during the time interval $\Delta t_{n}$ is modeled as uniformly accelerated motion as follows:
\begin{equation}
    \Delta {{r}_{n}}={{\omega }_{n}}+\frac{1}{2}{{\alpha }_{n}}\Delta t_{n}^{2}
    \label{eq37}
\end{equation}
where $\Delta r_{n}$ is the angle the MU$_{l}$ moved during the time interval $\Delta {{t}_{n}}$. $\Delta {{t}_{n}}$ means the time interval for beam reconstruction between time index $t_{n}$ and $t_{n+1}$. ${{\omega }_{n}}$ is the angular velocity of the MU$_{l}$. In the 3D beamforming scenario, two angular velocities ${{\omega }_{{{u}_{n}}}}$ and ${{\omega }_{{{v}_{n}}}}$ can be calculated based on the prediction angles at time $t_{n}$ and $t_{n-1}$ as follows: 
\begin{equation}
    \text{ }{{\omega }_{{{u}_{n}}}}=\frac{{{u}_{n+1}}-{{u}_{n}}}{\Delta {{t}_{n}}}
    \label{eq38}
\end{equation}
\begin{equation}
    \text{ }{{\omega }_{{{v}_{n}}}}=\frac{{{v}_{n+1}}-{{v}_{n}}}{\Delta {{t}_{n}}}
    \label{eq39}
\end{equation}
where $u_{n}$, $u_{n+1}$, $v_n$, $v_{n+1}$ are the spatial angles from MU$_l$ to U-UAV$_i$ predicted by the trained GPR model at time $t_n$ and $t_{n+1} = t_n + \Delta {{t}_{n}}$. If $n=0$, the $u_n$, $u_{n-1}$, $v_n$, $v_{n-1}$ are the real-time spatial angles calculated by the collected locations of MU$_l$. Meanwhile, angular acceleration of the MU$_{l}$ ${{\alpha }_{{{u}_{n}}}}, {{\alpha }_{{{v}_{n}}}}$ can be calculated based on angular velocities as follows:
\begin{equation}
    {{\alpha }_{{{u}_{n}}}}=\frac{{{\omega }_{{{u}_{n+1}}}}-{{\omega }_{{{u}_{n}}}}}{\Delta {{t}_{n}}}
    \label{eq40}
\end{equation}
\begin{equation}
    {{\alpha }_{{{v}_{n}}}}=\frac{{{\omega }_{{{v}_{n+1}}}}-{{\omega }_{{{v}_{n}}}}}{\Delta {{t}_{n}}}
    \label{eq41}
\end{equation}
Thus, the azimuth and elevation angular velocities and acceleration of MU$_l$ can be calculated. Then, for the two different spatial angles, set $\Delta r_{n}$ in Eq. (\ref{eq37}) as HPBW ${{\theta }_{u}}$ and elevation plane HPBW ${{\theta }_{v}}$ of the beam, respectively, and take Eq. (\ref{eq16}) and Eq. (\ref{eq17}) into Eq. (\ref{eq37}). The optimal value $\Delta {{t}_{n}}\text{=}\Delta t_{n}^{*}$ can be calculated as follows:
\begin{equation}
\begin{split}
    \Delta t_{n}^{*}( {{u}_{n}},{{v}_{n}},{{\omega }_{{u}_{n}}},{{\omega }_{{v}_{n}}},{{\alpha }_{{u}_{n}}},{{\alpha }_{{v}_{n}}})=\min ( \Delta {{t}_{{u}_{n}}},\Delta {{t}_{{v}_{n}}}))
\end{split}
\label{eq42}
\end{equation}
where $\Delta {{t}_{{u}_{n}}}$ and $\Delta {{t}_{{v}_{n}}}$ can be calculated as follows:
\begin{equation}
\begin{split}
    \Delta {{t}_{{u}_{n}}}=\sqrt{2\times ( \frac{{{\theta }_{{u}_{n}}}}{2}-{{\omega }_{{u}_{n}}})/{{\alpha }_{{u}_{n}}}\ }
\end{split}
\label{eq43}
\end{equation}
\begin{equation}
\begin{split}
    \Delta {{t}_{{v}_{n}}}=\sqrt{2\times( \frac{{{\theta }_{{v}_{n}}}}{2}-{{\omega }_{{v}_{n}}})/{{\alpha }_{{v}_{n}}}\ }
\end{split}
\label{eq44}
\end{equation}
Because the azimuth and elevation HPBW of the constructed beam are limited. Thus, the value of $\Delta {{t}_{n}}$ is small when MU$_l$ moves fast, and vice versa. MU$_l$ can stay in the coverage area of the beam constructed by U-UAV, and the energy consumption for beam reconstruction can be decreased. The pseudo-code is listed in detail as below.
\begin{algorithm}
    \caption{Angle-aware beam-tracking with adaptive beam reconstruction}
    \label{TIAM}
    \renewcommand{\algorithmicrequire}{\textbf{Input:}}
    \renewcommand{\algorithmicensure}{\textbf{Output:}}
    
    \begin{algorithmic}
        \REQUIRE $u_{n-1}$, $v_{n-1}$, $u_{n}$, $v_{n}$ of MU$_{l}$, GPR models model$_{u}$, model$_{v}$ for $u$ and $v$ of MU$_{l}$, number of antennas $N_{UPA}=N_{x} \times N_{y}$, real-time $t_{s}$, time interval $\Delta t$  
        \ENSURE The optimized time interval $\Delta t_{n}^{*}$ for beam reconstruction   
        \WHILE{$t_{s}$}
        \IF {$t_{s}$ is time to collect data from the MU$_{l}$ and one of the errors between predictive angles $u$, $v$ and real angles $u$, $v$ is bigger than ${{\sigma }_{check}}$}
            \STATE Rebuild the GPR model for $u$ or $v$ whose predictive error is over ${{\sigma }_{check}}$ 
        \ENDIF
        \STATE Calculate the $\omega_{u_{n-1} } $, $\omega_{v_{n-1} } $, with $u_{n-1}$, $v_{n-1}$, $u_{n}$, $v_{n}$
        \STATE Calculate the HPBW with $N_x$, $N_y$, $u_{n}$, $v_{n}$
        \STATE Use model$_{u}$ and model$_{v}$ to predict $u_{n+1}$, $v_{n+1}$ at $T_{s}+\Delta t$
        \STATE Calculate the $\omega_{u_{n}}$, $\omega_{v_{n}}$ with $u_{n}$, $v_{n}$, $u_{n+1}$, $v_{n+1}$
        \STATE Calculate the angular acceleration with $\omega_{u_{n-1}}$, $\omega_{v_{n-1}}$, $\omega_{u_{n}}$, $\omega_{v_{n}}$
        \STATE Calculate the optimal $\Delta t_{n}^{*}$ with angular acceleration ${{\alpha }_{u}},{{\alpha }_{v}}$, HPBW and $\omega_{u_{n}}$, $\omega_{v_{n}}$
        \STATE Update $\Delta t$ as $\Delta t_{n}^{*}$
        \STATE Update $T_{s}$ as $T_{s}+\Delta t$
        \STATE Update $u_{n-1}$ as $u_{n}$, $v_{n-1}$ as $v_{n}$
        \STATE Update $u_{n}$, $v_{n}$ as model$_{u}$, model$_{v}$ predict at $t_{s}$
        \STATE Calculate the beam vector from the U-UAV$_{i}$ to the MU$_{l}$ with $u_{n}$, $v_{n}$
        \STATE Build beam from the U-UAV$_{i}$ to the MU$_{l}$ 
        \ENDWHILE
        \RETURN The optimal result $(\hat{x},\hat{y})$ as the location of U-UAV$_{i}$
    \end{algorithmic}
\end{algorithm}
\subsection{Theoretical Analysis of the Proposed BAB-AR}
Owing to building 3D beam tracking between UAV and MU, the computational complexity of the optimized beam vector algorithm is proportional to the number of antennas\cite{XChen2020}. Thus, the complexity of the optimized function is $\mathcal{O}$($N^{3.5}$), where $N$ is the number of the UPA's antennas \cite{LPZhu2020} \cite{ZYXiao2020}. Meanwhile, the optimal value of the beam vector is dependent on the number of iterations performed by the algorithm. Thus, the complexity of the beam vector-optimized method is $\mathcal{O}$($N_{iter} \texttimes N_{3.5}$), where $N_{iter}$ denotes the number of iterations required for convergence. 

In the proposed BAB-AR, a GPR algorithm is used to predict the angles of the target MU based on historical trajectory, which eliminates extra time for channel estimation in the 3D beam-tracking scenario. In addition, the trajectory information collection for the MU and U-UAVs localization are parallel process. Thus, the GPR model training can be completed once the U-UAV is located. The time required for model training can be ignored as the GPR model can be trained with minimal data in a short period. The total time consumption of BAB-AR is only dependent on the time taken by A-UAVs to locate U-UAVs using GDCSA. Since GDCSA's computation dimension is always two, the complexity of solving Eq. ($\ref{eq25}$) is only relative to the dimension of $a_{i,k}$, which is $N_{ULA}$, and the complexity of the proposed BAB-AR is $\mathcal{O}$($N_{ULA} \times N_{iter} \times N_{pop}$), where $N_{pop}$ is the number of the initial populations.

The hybrid-codebook beam tracking algorithm has the lowest complexity $\mathcal{O}$($N_c$) among the aforementioned methods. The complexity of hybrid-codebook beam tracking algorithm is only dependent on the number of the fixed codebook $N_c$. However, it fails to provide high beam gain for MU$_l$.
\section{Experiments and discussions}
This section verifies the robustness of the proposed BAB-AR algorithm against three baseline algorithms: hybrid codebook-based beam-tracking \cite{AAlkhateeb2014}, beam vector-optimized beam-tracking algorithm \cite{XChen2020}, and angle-aware beam-tracking algorithm without data processing based on GPR \cite{HLSong2021}. This study considers UAV-BSs with a height of 100m distributed in an area of  $[(0,0), (150,150)]$. The number of antennas for U-UAVs is set to $16\times 16$, while the number of antennas for A-UAVs is set to 16. During the experiments, $k$ is set to 2, and the coordinates of A-UAVs are (0, 0), (50, 50), respectively. In the phase of collecting MU$_l$'s trajectory, A-UAVs collect location information from the MU$_l$ every $\Delta t_{n}$. The prediction model is rebuilt at each $t_{check}$. The meaning of the variables used in this section is listed in Table\ref{table2}.
\begin{table*}[t!]
\centering
\vspace{-2.0em}
\caption{Variables in experiments}
\label{table2}
\begin{tabular}{|c||c||c|}
\hline
\textbf{Parameter} & \textbf{Value} & \text{Description} \\ \hline
$h$ & $100$ & Drone flight altitude \\  
\hline
$N_{x}\times N_{y}$& $16\times 16$& Specifications of the U-UAV equipped with  a surface array\\ \hline
$N_{ULA}$& $16$& Specifications for drones equipped  with line arrays\\
\hline
$k$ & $2$& Number of A-UAVs\\
\hline
$(x_{i},y_{i})$& $(x_{i},y_{i})$ &Location of the U-UAV$_{i}$ within the flight area\\
\hline
$u_{i}$& $u_{i}$ &The angle of U-UAV$_{i}$ calculated based on $(x_{i},y_{i})$ within the flight area\\
\hline
$\theta _{i,k}$ & $\theta _{i,k}$ &The spatial angle from the U-UAV$_{i}$  to the A-UAV$_{k}$\\
\hline
$(u_{l},v_{l})$ & $(u_{l},v_{l})$ & The spatial angle from the MU$_{l}$ to the U-UAV$_{i}$\\
\hline
$({{\hat{x}}_{i}},{{\hat{y}}_{i}})$& $({{\hat{x}}_{i}},{{\hat{y}}_{i}})$ & The predicted location of the resulting the U-UAV$_{i}$ within the flight area\\
\hline
${\hat{u}}_{i}$& ${\hat{u}}_{i}$ & The angle of U-UAV$_i$ calculated based on $({{\hat{x}}_{i}},{{\hat{y}}_{i}})$ within the flight area\\
\hline
${{\hat{\theta }}_{i,k}}$ & ${{\hat{\theta }}_{i,k}}$ & The predicted spatial angle from the U-UAV$_{i}$ to the A-UAV$_{k}$\\
\hline
$({{\hat{u}}_{l}},{{\hat{v}}_{l}})$&$({{\hat{u}}_{l}},{{\hat{v}}_{l}})$ & The predicted spatial angle from the MU$_{l}$ to the U-UAV$_{i}$\\
\hline
$n$ &\thead{when $\Delta t_{n}=0.1s$, $n\in [0,50]$; \\ when $\Delta t_{n}=0.01s$, $n\in [0,500]$;}& Number of the time index in a trajectory\\
\hline
$t_{n}$& $n\times \Delta t_{n}$ & Time index during the experiment\\
\hline
$\Delta t_{n}$ & 0.1s& Time interval for the the UAV$_{i}$ to collect the location information of the MU$_{l}$\\
\hline
$\Delta t_{n}^{*}$ & Dynamical time interval and the initial value is 0.1s & Time interval for predicting the spatial angle of the MU$_{l}$\\
\hline
$t_{check}$ & 3s & Time interval for checking if GPR model should be rebuilt\\
\hline
${{\sigma }_{check}}$ &0.05$\pi$ &Threshold value for checking if GPR model should be rebuilt\\ 
\hline
\end{tabular}
\end{table*}
\subsection{Accuracy analysis of angle prediction with BAB-AR algorithm}
\subsubsection{Angle prediction of UAV-BSs}
The proposed GDCSA is used by A-UAVs to locate U-UAVs in a cooperative manner. Table \ref{table3} demonstrates that the optimized results obtained by three distinct algorithms. The proposed GDCSA exhibits superior accuracy over both CSA and PSO algorithms. Furthermore, the GDCSA algorithm requires less running time to obtain the optimal result in comparison to the original CSA algorithm.
\begin{table}
    \caption{Comparison of the convergence speed}
    \label{table3}
	\centering
	\begin{tabular}{|c||c||c|}
	\hline
 Algorithm &Value of fitness & Running time/s{} \\ \hline
GDCSA & 2.81& 1.09\\
CSA & 10.95& 1.38 \\
PSO& 80.84 & 1.13 \\ 
\hline
\end{tabular}
\end{table}
\begin{figure}[H]
\centering
\subfloat[CTRV.]{\includegraphics[width=.15\textwidth]{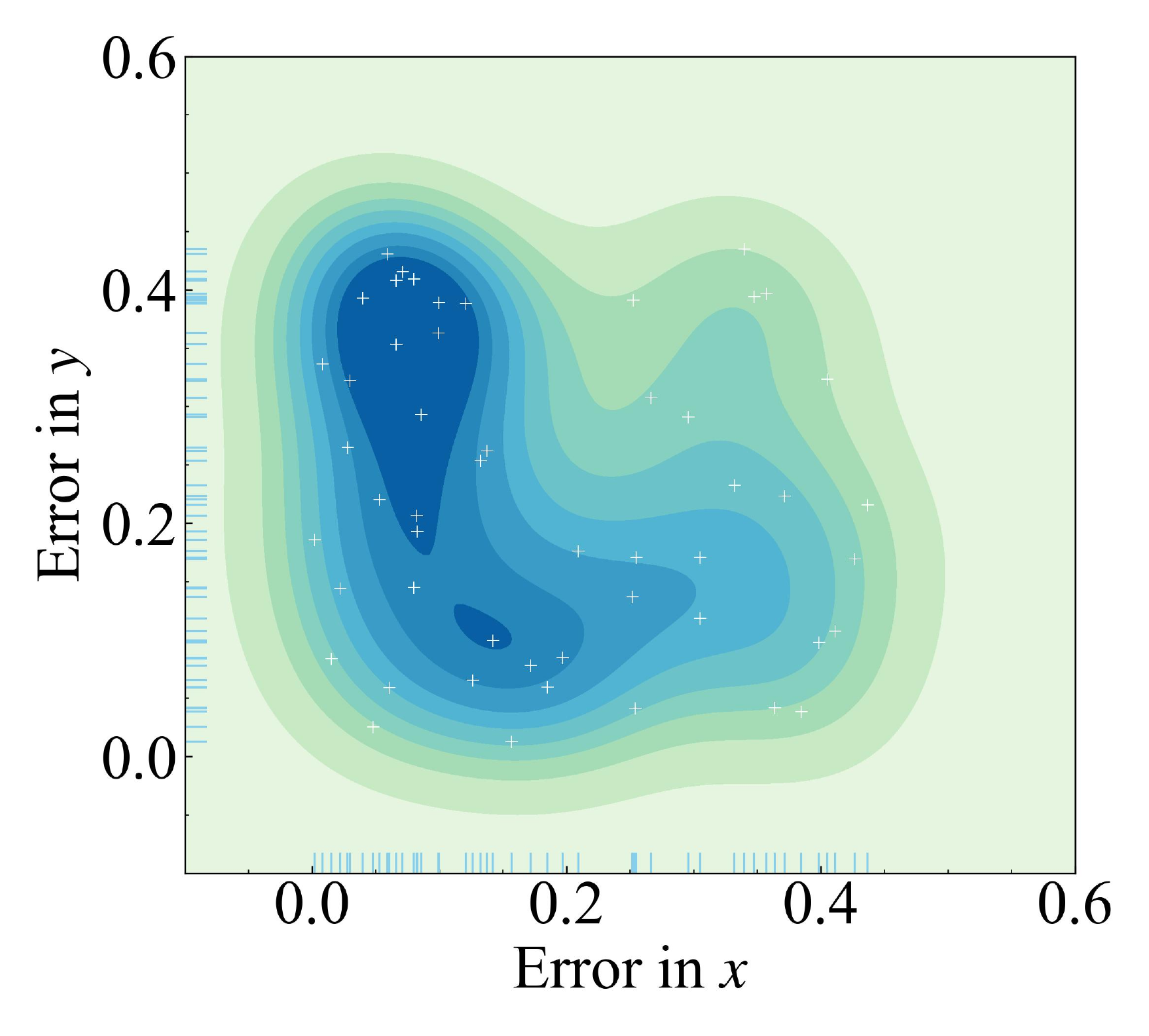}%
\label{fig7a}}
\hfil
\subfloat[CTRA.]{\includegraphics[width=.15\textwidth]{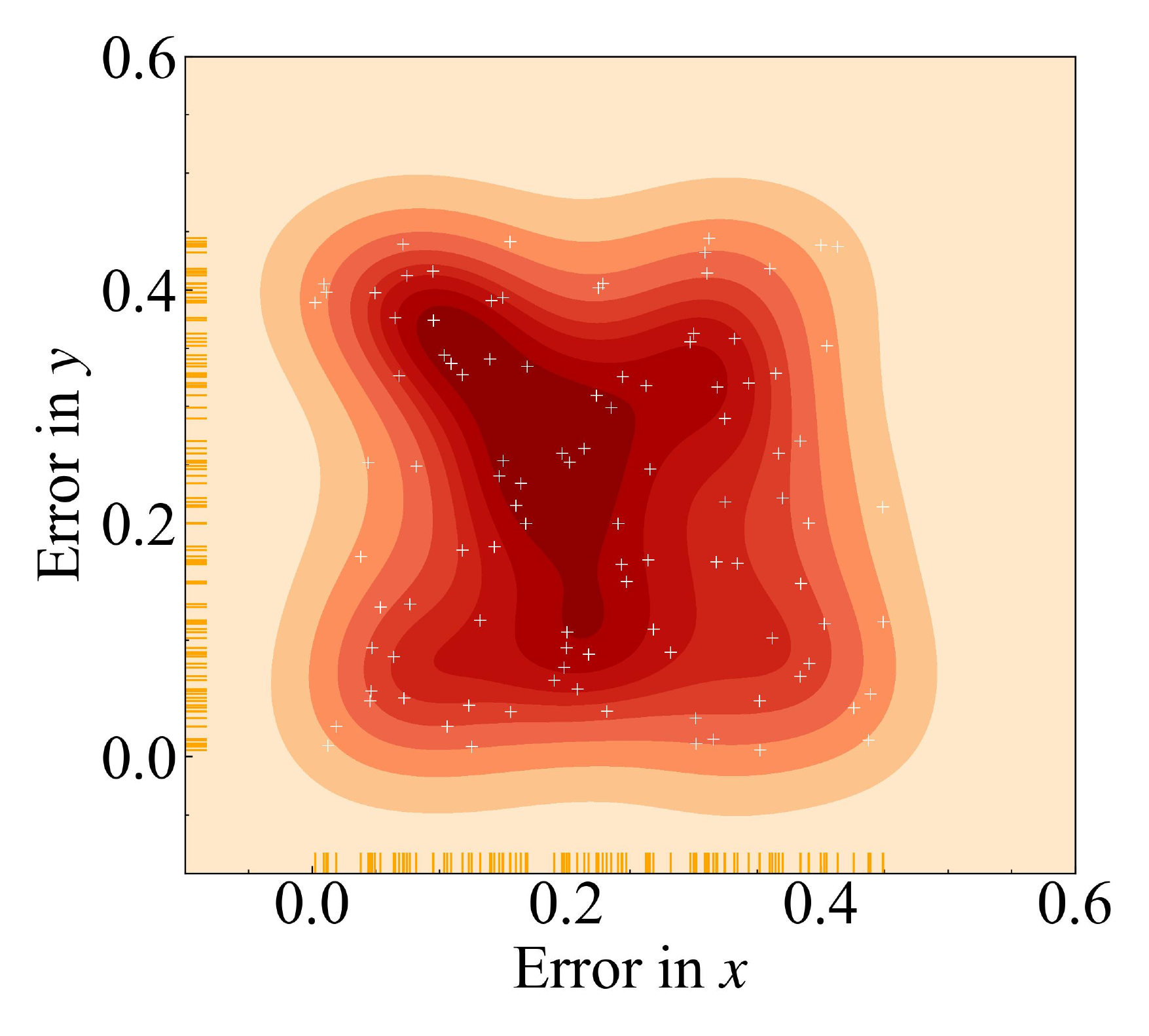}%
\label{fig7b}}
\hfil
\subfloat[Random.]{\includegraphics[width=.15\textwidth]{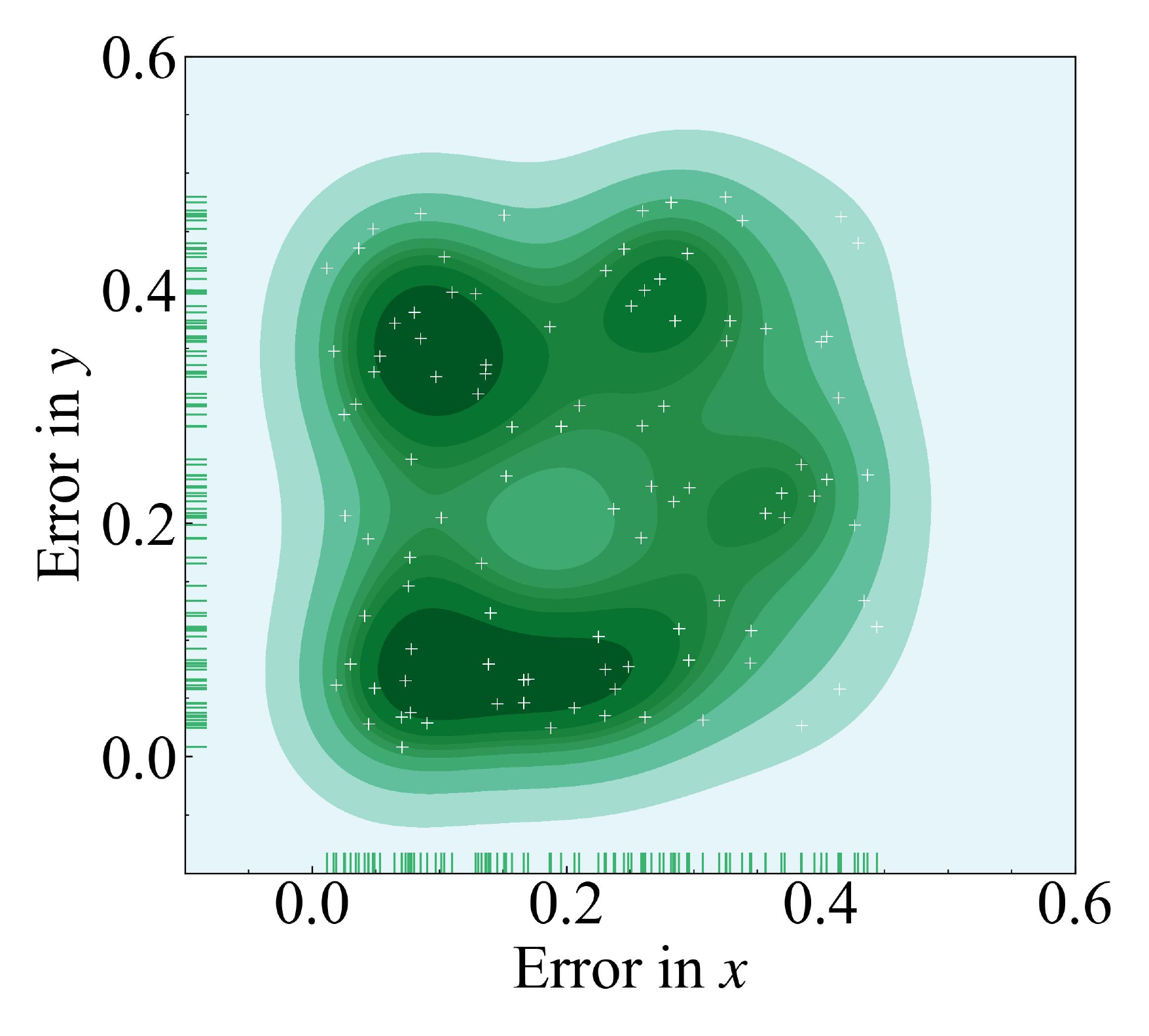}%
\label{fig7c}}
\caption{Error between $({{\hat{x}}_{i}},{{\hat{y}}_{i}})$ and $(x_{i},y_{i})$ in different moving patterns.}
\label{fig7}
\end{figure}
Fig. \ref{fig7} shows the prediction error of the target U-UAV from time index \textit{$t_0$} to \textit{$t_{50}$} in different moving patterns, including constant turn rate and velocity (CTRV), constant turn rate and acceleration (CTRA), and random wandering. The colors of the regions represent the prediction error distribution between $({{\hat{x}}_{i}},{{\hat{y}}_{i}})$ predicted by GDCSA and real $(x_i, y_i)$, and the boundary of the regions represent the range of error values. Specifically, the darker color of the region represents the denser distribution of error values. As can be seen, the proposed GDCSA can precisely locate U-UAVs while the errors between $({{\hat{x}}_{i}},{{\hat{y}}_{i}})$ and $(x_i, y_i)$ are all within 0.5 m in different moving patterns, which means that the performance of GDCSA is stable in different moving patterns. The reason is that the localization process in GDCSA is not dependent on the trajectory of U-UAVs, but rather on the strength of the reference signal received at A-UAVs.

Based on the optimized locations, the spatial angles from the U-UAV$_{i}$ to the A-UAV$_{k}$ can be calculated. Fig. 8 shows the relative errors between $u_i$ and $\hat{u_i}$ estimated by GDCSA and the GPR algorithm proposed in \cite{HLSong2021}. CTRV, CTRA, and random wandering moving patterns are considered in this experiment. In Fig. 8(a), the UAV moves without any shaking, while in Fig. 8(b), the UAV’s locations are fluctuated within a 50m radius by the UAV shaking. As shown in Fig. 8(a), the angle relative error of the BAB-AR algorithm is less than 0.2 under different motion patterns. However, the angle relative error of the algorithm \cite{HLSong2021} is less than 0.2 only when the UAVs move under the CTRV moving pattern. In random moving pattern, its relative error is 0.6. In brief, the GDCSA can predict spatial angles with high accuracy under various UAV moving patterns. Meanwhile, Fig. 8(b) shows that the relative error in predictive angle calculated by GPR has obvious fluctuation due to the influence of shaking, as the historical trajectory information is used for spatial angle prediction. However, the accuracy of the predictive angle which is calculated by the predictive location cannot be affected by UAV shaking, and the error caused by GDCSA maintains a value less than 0.2.
\begin{figure}
\centering
\subfloat[UAV with no shaking.]{\includegraphics[width=1.5in,height=1.25in]{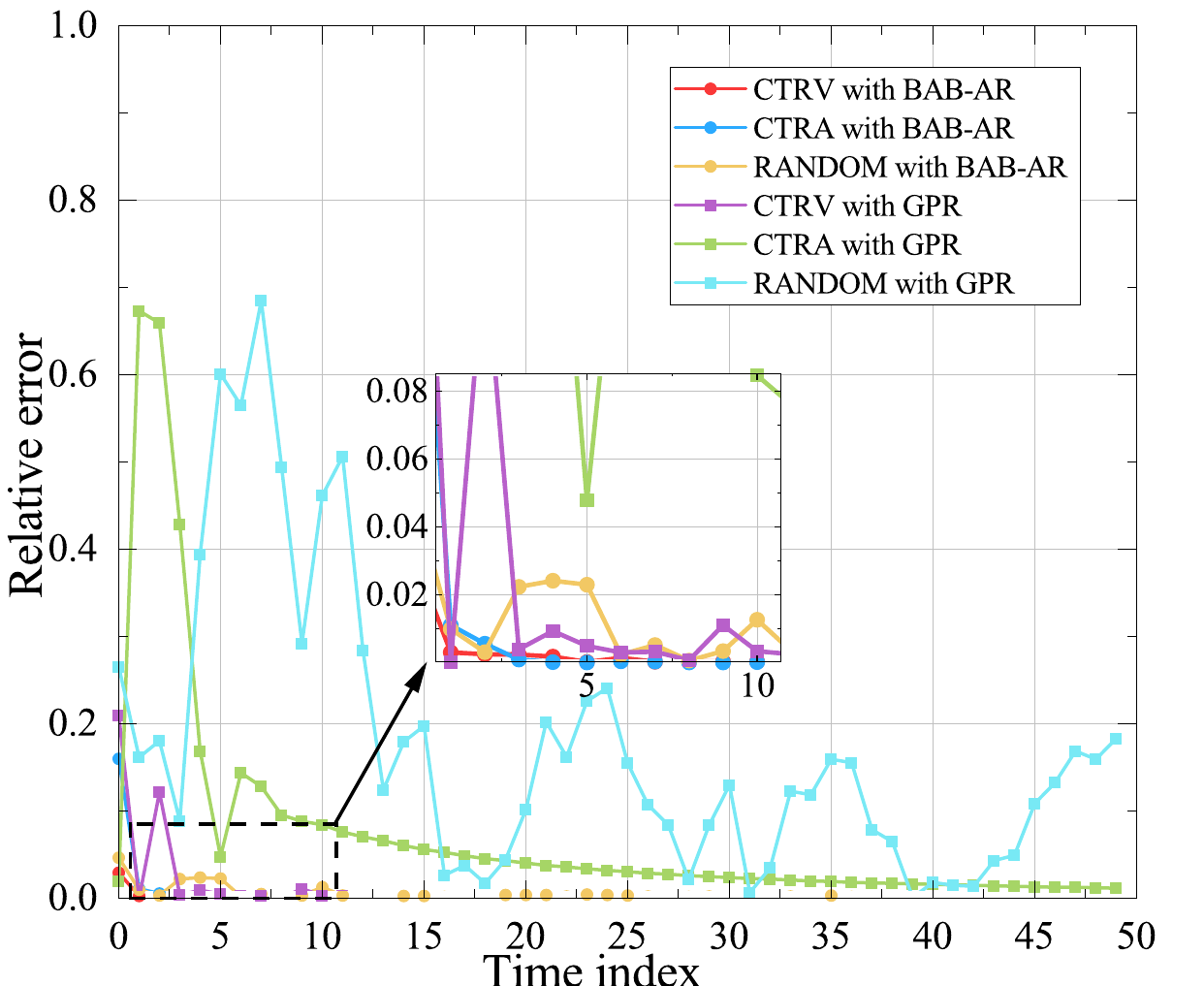}%
\label{fig8a}}
\hfil
\subfloat[UAV with shaking.]{\includegraphics[width=1.5in,height=1.25in]{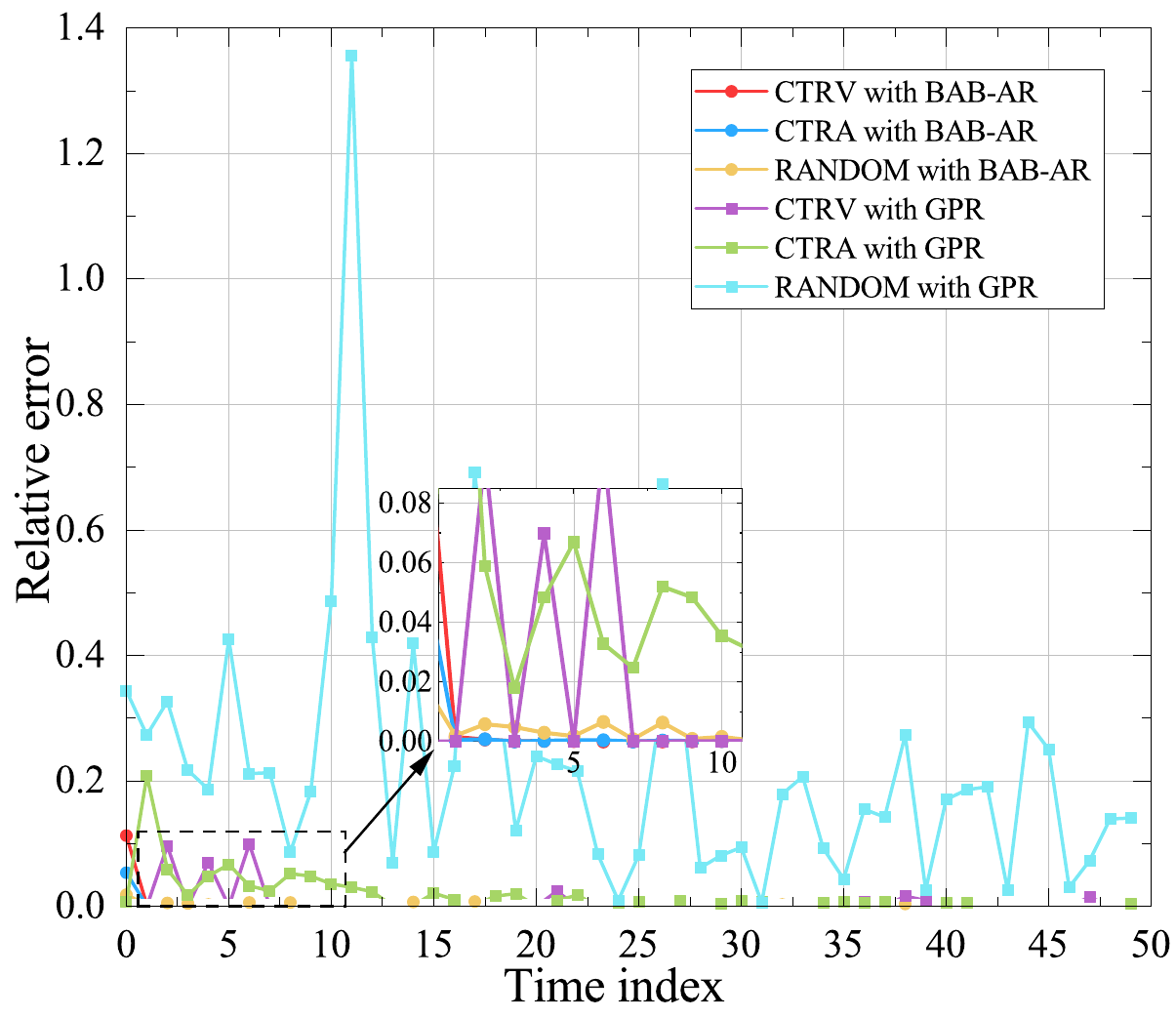}%
\label{fig8b}}
\caption{Relative errors between ${{\theta }_{i,k}}$, and ${{\hat{\theta }}_{i,k}}$ in different moving patterns.}
\label{fig8}
\end{figure}
\par
\begin{figure}
    \centering
    \includegraphics[width=2.5in,height=1.43in]{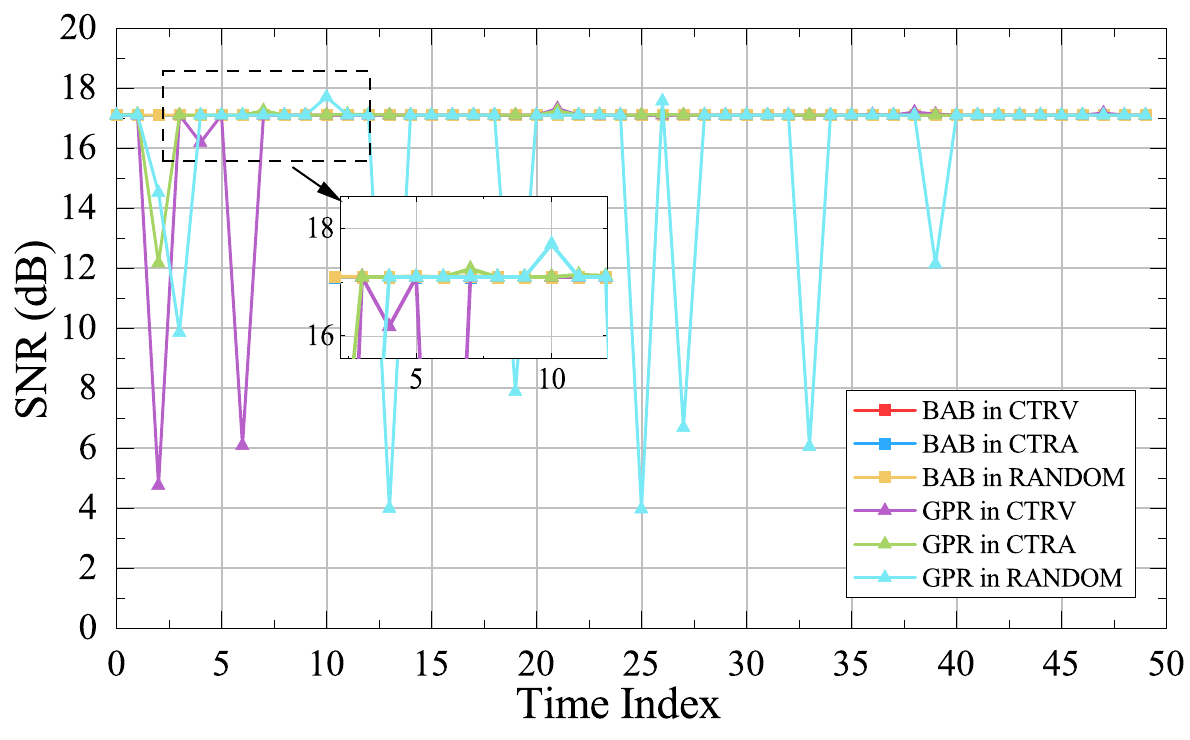}
    \caption{SNR under different moving patterns.}
    \label{fig9}
\end{figure}

 In addition, the SNR at different U-UAVs under different moving patterns is also used to evaluate the performance of GDCSA. The SNR of the signal received at U-UAV$_i$ can be calculated as follows:
\begin{equation}
   \begin{split}
   SNR=\frac{{{p}_{k,i}}{{\left| {{r}_{k,i}} \right|}^{2}}}{{{\sigma }^{2}}}
   \label{eq45}
 \end{split}
\end{equation}
where $p_{k, i}$ is the transit power from A-UAV$_k$ to U-UAV$_i$, which is set as 20 dBm and the noise power is set as -100 dBm. Meanwhile, ${r}_{k,i}$ is the signal received at U-UAV$_i$ from A-UAV$_k$ as expressed in Eq. (\ref{eq6}). As shown in Fig. \ref{fig9}, the shaking of UAVs does not affect the beam alignment, and high SNR can be maintained under different moving patterns by GDCSA. Because the location prediction of the proposed GDCSA is calculated based on the received signal strength of A-UAV. However, the angle prediction errors affect the beam alignment among U-UAVs in GPR. 
\par
\begin{figure}
    \centering
    \includegraphics[width=2.5in,height=1.43in]{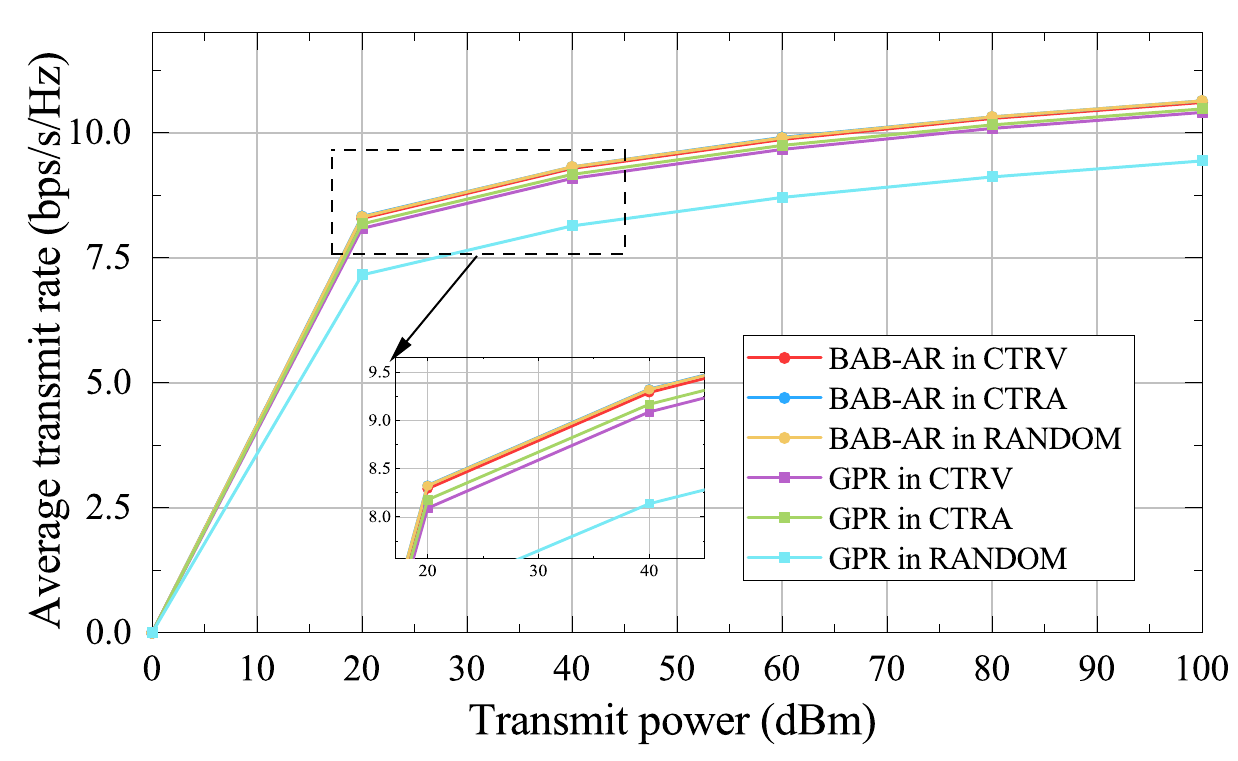}
    \caption{Average rate under different transmit power.}
    \label{fig10}
\end{figure}

 Besides, the transmission rate, which can be expressed as Eq. (\ref{eq46}) can also be improved with a high SNR:
\begin{equation}
   \begin{split}
   Rate={{\log }_{2}}\left( 1+SNR \right)
   \label{eq46}
 \end{split}
\end{equation} 
Fig. \ref{fig10} shows that the average data transmission rate at U-UAVs with GDCSA is capable of achieving data transmission among UAV-BSs with high efficiency.
\subsubsection{Angle prediction of the target MU}
\begin{figure}
\centering
\subfloat[BAB-AR in CTRV moving pattern.]{\includegraphics[width=1.5in,height=1.25in]{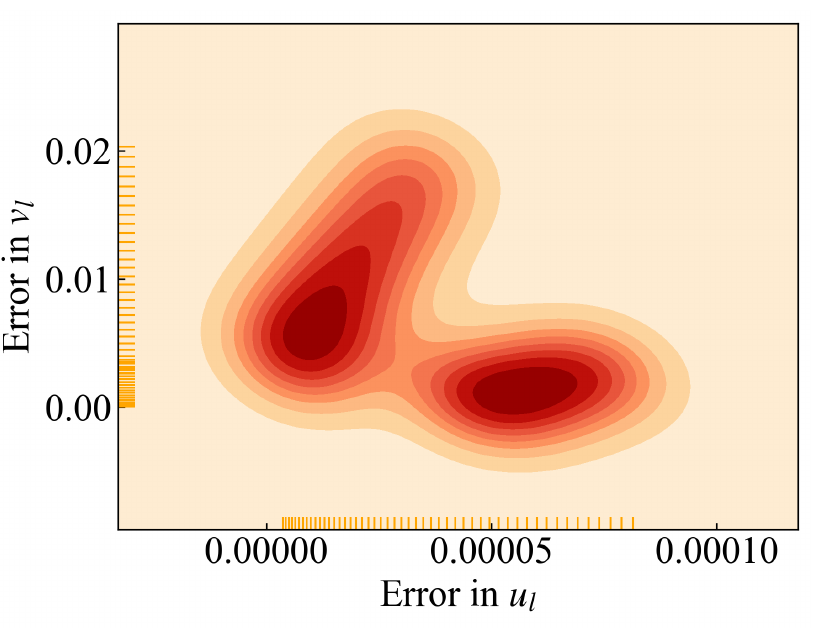}%
\label{fig11a}}
\hfil
\subfloat[GPR in CTRV moving pattern.]{\includegraphics[width=1.5in,height=1.25in]{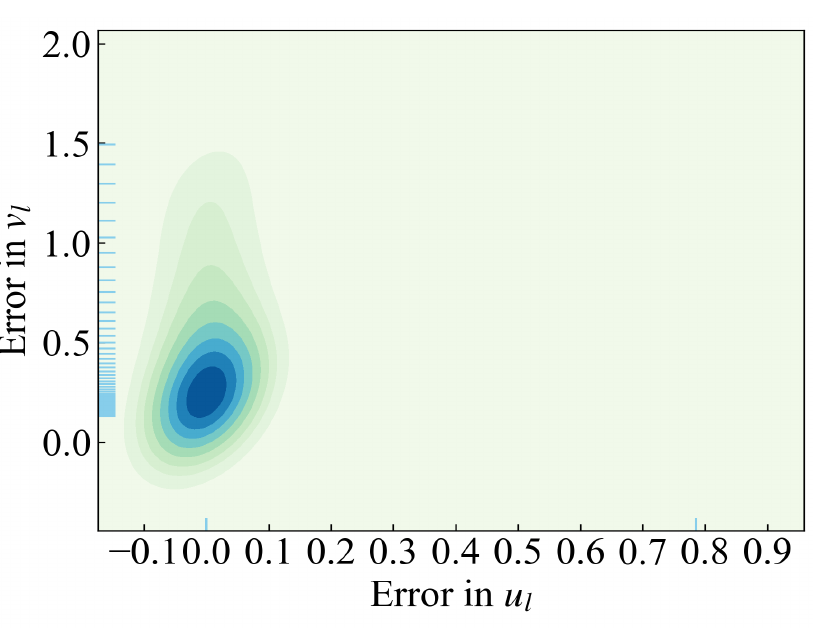}%
\label{fig11b}}
\hfil
\subfloat[BAB-AR in CTRA moving pattern.]{\includegraphics[width=1.5in,height=1.25in]{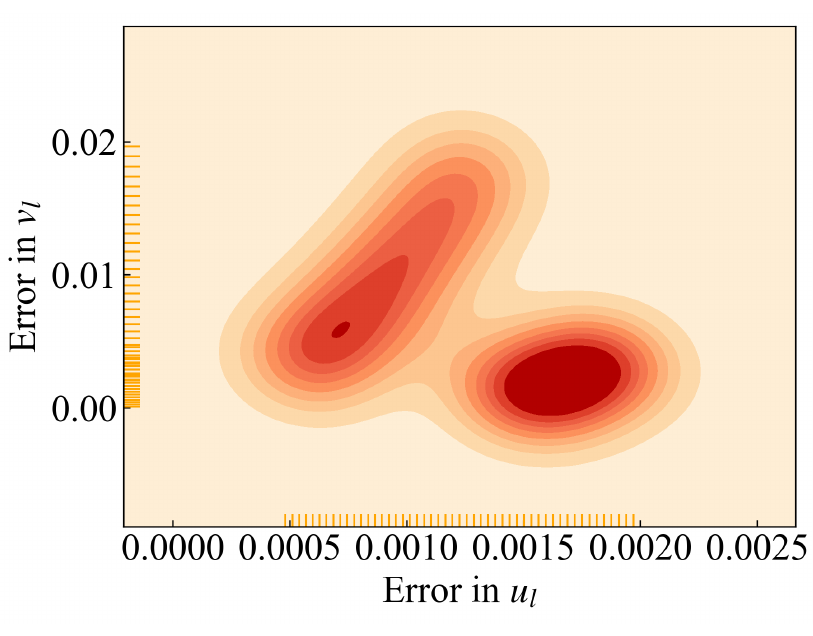}%
\label{fig11c}}
\hfil
\subfloat[GPR in CTRA moving pattern.]{\includegraphics[width=1.5in,height=1.25in]{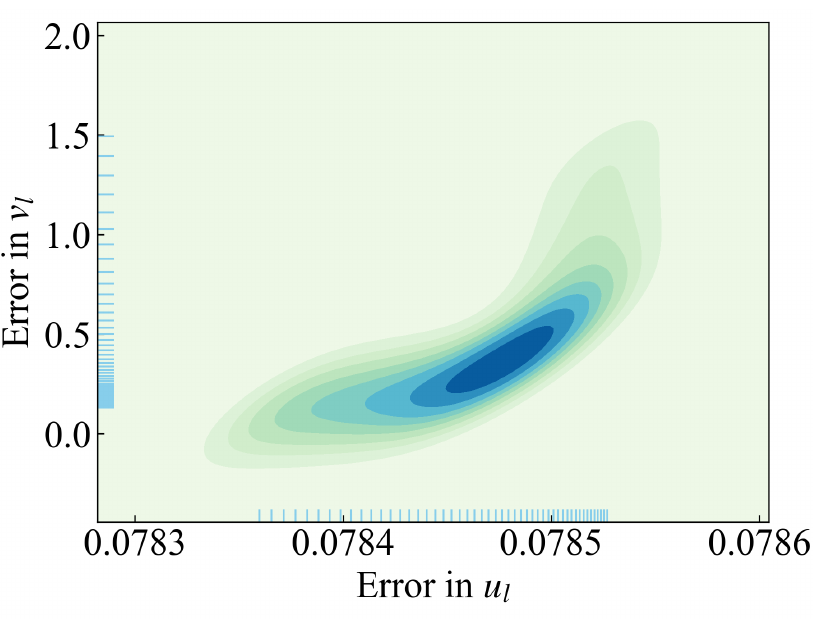}%
\label{fig11d}}
\caption{Errors between $(u_l, v_l)$ and in the CTRA moving pattern.}
\label{fig11}
\end{figure}
\par During the beam tracking phase between UAV and MUs, a trained GPR model is used to predict the spatial angle of the target MU$_{l}$. As illustrated in Fig. \ref{fig11}, predictive errors between $(u_{l}, v_{l})$ and $({{\hat{u}}_{l}},{{\hat{v}}_{l}})$ are presented under CTRV and CTRA moving patterns with $\Delta t=0.1s$, respectively. The GPR without data processing exhibits instability. Nevertheless, the BAB-AR algorithm demonstrates an error margin of 0.025 or less under each moving pattern. The high accuracy of  spatial angles prediction ensures the maintenance of high beam gain, as discussed in Section V.B.
\begin{figure}
\centering
\subfloat[Azimuth angle $u$ from MU$_l$ to U-UAV$_i$. ]{\includegraphics[width=1.5in,height=1.25in]{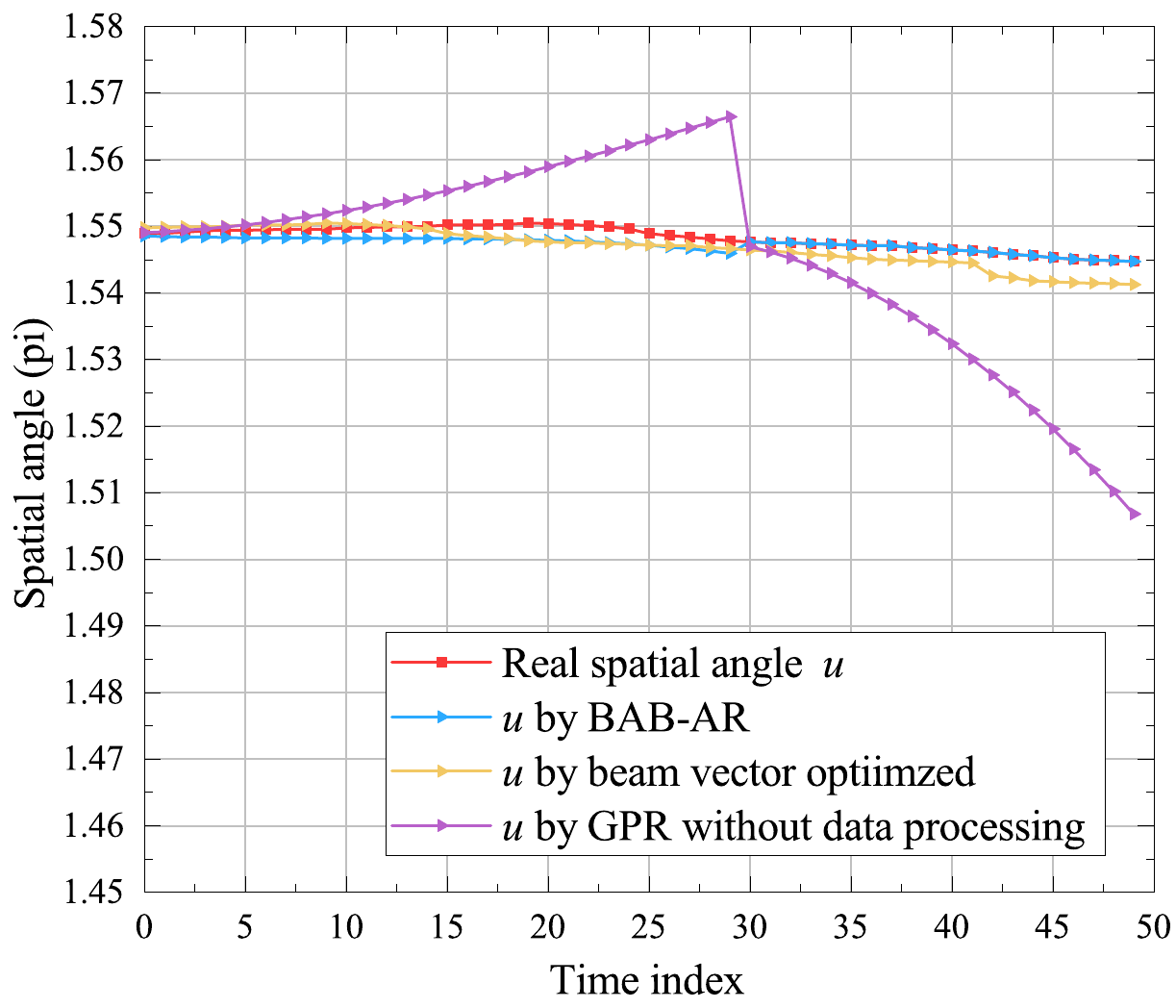}%
\label{fig12a}}
\hfil
\subfloat[Elevation angle $v$ from MU$_l$ to U-UAV$_i$.]{\includegraphics[width=1.5in,height=1.25in]{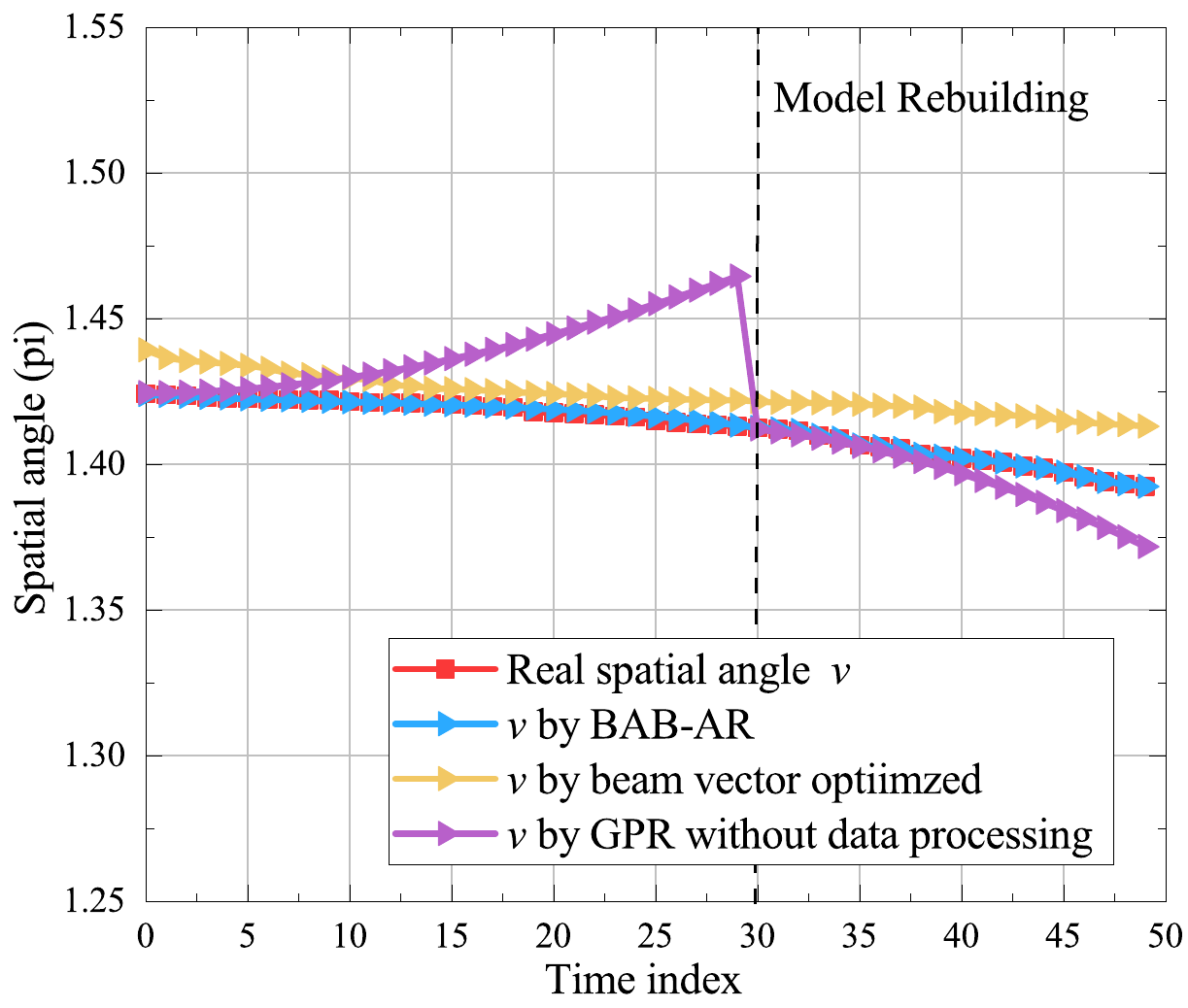}%
\label{fig12b}}
\caption{Angle prediction using different algorithms.}
\label{fig12}
\end{figure}
In Fig. \ref{fig12}, the real vehicle trajectory data\footnote {The simulation data of trajectory of GU is from “https://data.transportati-\\on.gov”} is used to verify the accuracy of the BAB-AR algorithm by comparing with other algorithms. 
The vehicle trajectory data was collected by Federal Highway Administration from a specified freeway segments in the USA. As shown in Fig. \ref{fig12}, BAB-AR can accurately predict the angles of MU. The error between $u_l$ and $\hat{u_l}$ estimated by BAB-AR, beam vector optimized algorithm \cite{XChen2020} and the angle predictive beam tracking algorithm based on GPR \cite{HLSong2021} are 0.1146°, 0.5675°, and 0.0114°, respectively. Similarly, the error between $v_l$ and $\hat{v_l}$ are 0.1547°, 0.6535°, and 0.6477°, respectively. The proposed BAB-AR algorithm has Superior accuracy than the GPR algorithm without data processing \cite{HLSong2021}. However, the alignment angle of the beam vector-optimized algorithm \cite{XChen2020} satisfies accuracy requirements only when the spatial angle changes slowly as shown in Fig. 12(a). Because the beam tracking between UAV and MU is 3D beam-tracking channel, which takes more time for the beam vector-optimized algorithm \cite{XChen2020} to compute the optimal beam vector. \par In BAB-AR, the trajectory used for training the GPR model should adapt to the movement of the MU$_l$ in time. In this experiment, the update frequency of the GPR model $t_{check}$ is set to 3s, and the error threshold ${{\sigma }_{check}}$ between predictions and real data is set to 0.05$\pi$. In Fig. \ref{fig12}, the BAB-AR model is not rebuilt for azimuth angle $u$, but the model for elevation angle $v$ is rebuilt at time index 30.
\subsection{Comparison of beam gain at MU}
\begin{figure}
\centering
\subfloat[Beam gain with BAB-AR at $t_n$=0.]{\includegraphics[width=1.5in,height=1.25in]{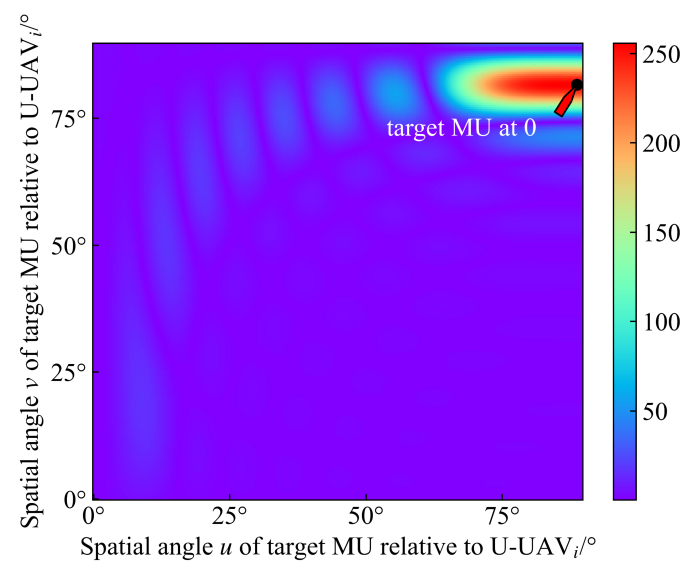}%
\label{fig13a}}
\hfil
\subfloat[Beam gain with hybrid codebook-based at $t_n$=0.]{\includegraphics[width=1.5in,height=1.25in]{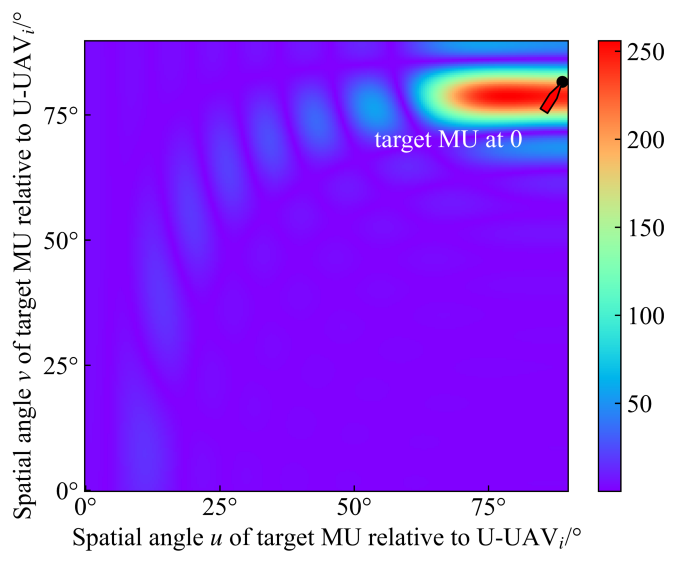}%
\label{fig13b}}
\hfil
\subfloat[Beam gain with BAB-AR at $t_n$=50.]{\includegraphics[width=1.5in,height=1.25in]{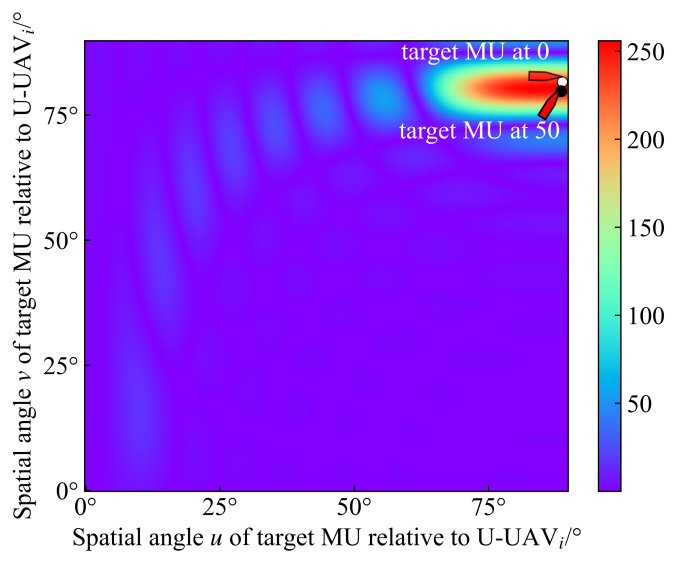}%
\label{fig13c}}
\hfil
\subfloat[Beam gain with hybrid codebook-based at $t_n$=50.]{\includegraphics[width=1.5in,height=1.25in]{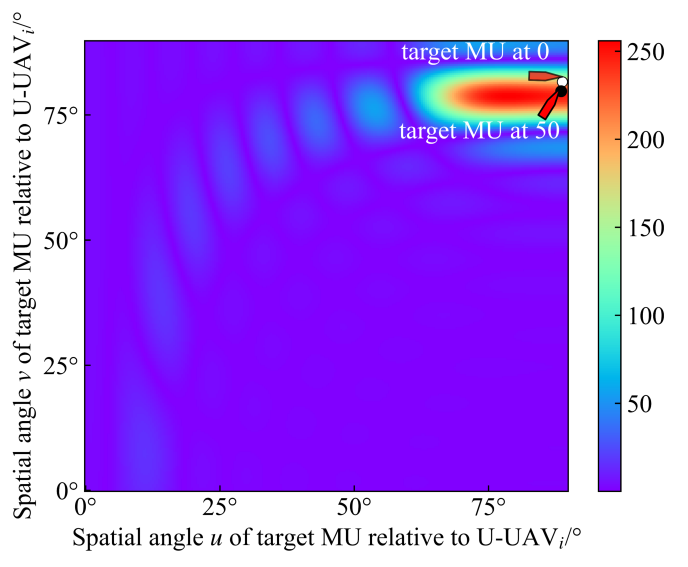}%
\label{fig13d}}
\caption{Beam gain heat map.}
\label{fig13}
\end{figure}

In this experiment, the beam gain at the target MU can be calculated as follows:
\begin{equation}
    {{G}_{GU}}=\sqrt{{{N}_{x}}\times {{N}_{y}}}{{w}_{l}}\cdot H\cdot a_{l}^{H}
    \label{eq47}
\end{equation}
The beam gain of the hybrid codebook-based beam-tracking algorithm \cite{AAlkhateeb2014} and BAB-AR algorithm at $t_n=0$ and $t_n=50$ are shown in Fig. \ref{fig13}, respectively. The maximum beam gain region is highlighted in red, while regions with decreasing beam gain are colored green and purple. At $t_n=0$, the MU$_l$ is located at the edge of the beam constructed by the hybrid codebook-based approach, which results in a green-colored region with a beam gain value of 172.57. At $t_n=50$, the MU$_l$ has moved to a region aligned with one of the constructed codebooks, which causes the color of region to change to red, and the beam gain value is increased to 234.78. In contrast, for BAB-AR, regions surrounding the MU$_l$ are colored red at the $t_n=0$ and $t_n=50$, with beam gain value of 255.88 and 255.65, respectively. Thus, BAB-AR can construct an optimal beam for the MU$_l$ promptly.
\par
\begin{figure}
    \centering
    \includegraphics[width=2.5in,height=1.43in]{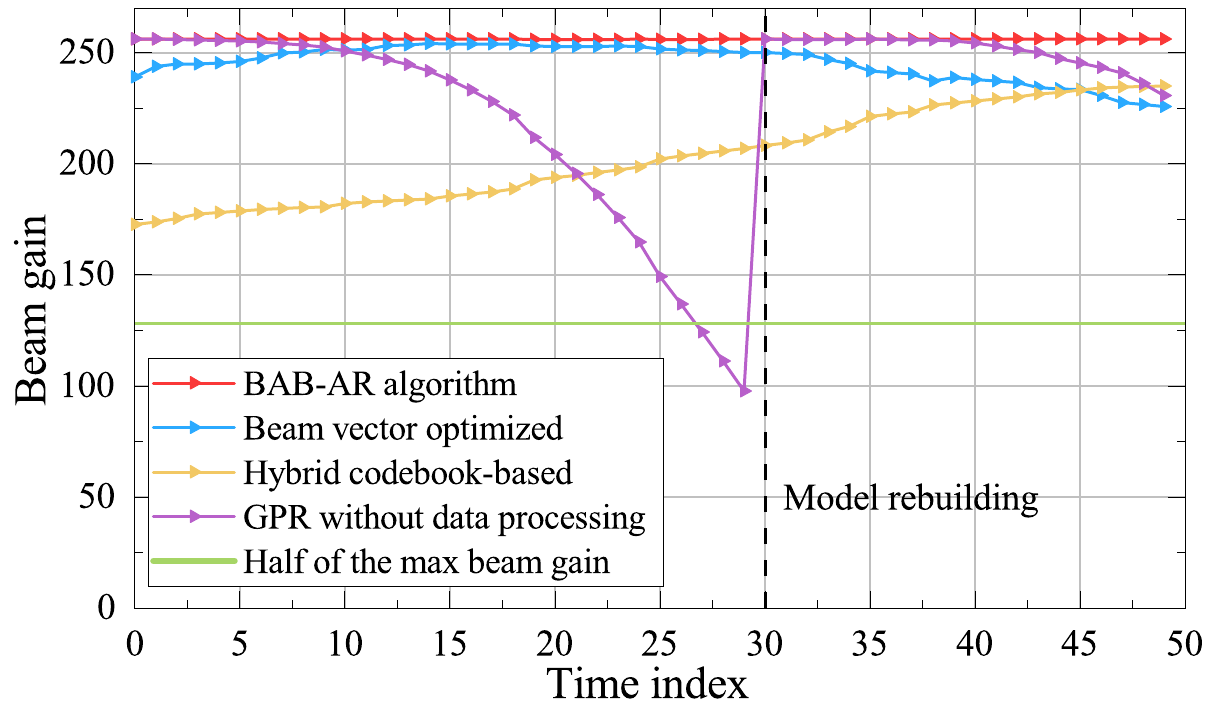}
    \caption{Beam gain at the MU$_l$ per $\Delta t=0.1s$ with different algorithms.}
    \label{fig14}
\end{figure}

 The beam gain of different algorithms with the real trajectory of MU is simulated as shown in Fig. \ref{fig14}. At $t_{check}=3s$, the GPR model is rebuilt when $t_{n}=30$. The beam gain of \cite{XChen2020} is lower than that of other algorithms as the trajectory changes rapidly after $t_n=30$. With the beam-tracking algorithm \cite{AAlkhateeb2014}, the beam gain remains substantial only when the MU$_l$’s location aligns with the coverage area of the constructed beam. The predictive accuracy of the GPR algorithm without data processing is inferior to that of the BAB-AR algorithm. The beam gain of the BAB-AR algorithm received at the MU$_l$ is improved by approximately 5.52\%, 21.58\%, and 4.51\% compared to those achieved by the beam vector optimized algorithm \cite{XChen2020}, hybrid codebook-based beam-tracking algorithm \cite{AAlkhateeb2014}, and the angle-predictive beam tracking realized by GPR without data processing \cite{HLSong2021}. 
\subsection{Importance of dynamic beam reconstruction}
\begin{figure}
    \centering    \includegraphics[width=2.5in,height=1.43in]{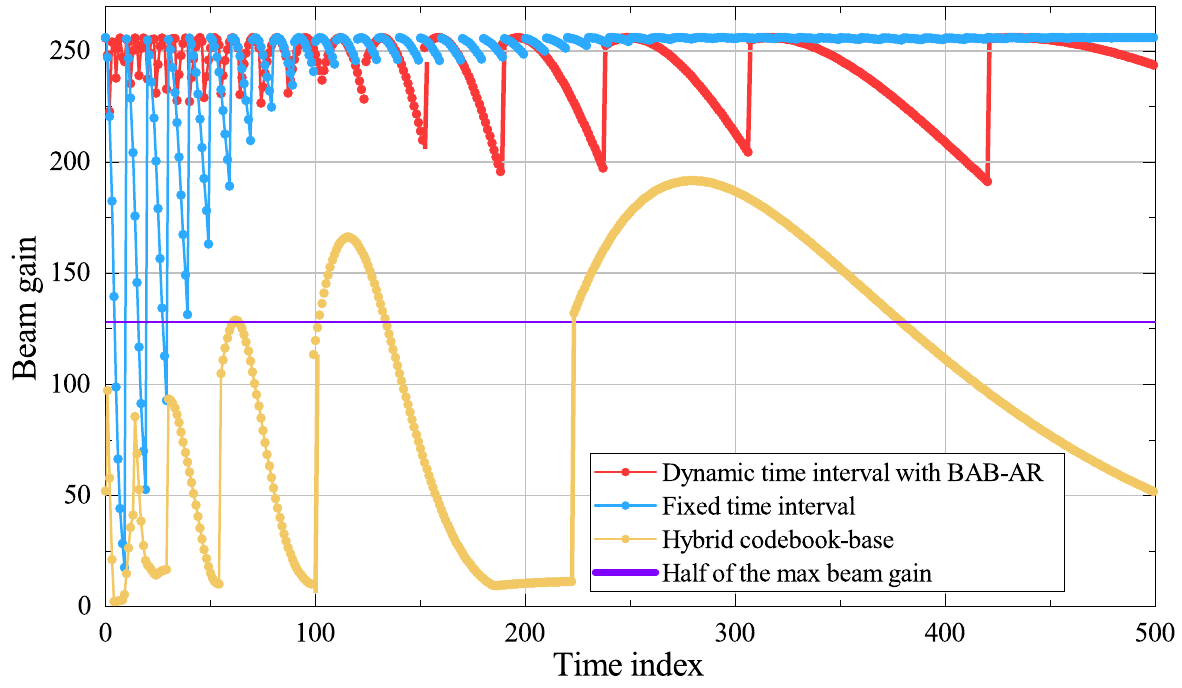}
    \caption{Beam gain received at the MU$_l$ per $\Delta t=0.01$s  with different algorithms.}
    \label{fig15}
\end{figure}

The importance of dynamic beam reconstruction is shown in Fig. \ref{fig15}. A CTRA moving pattern with a time interval of $\Delta t=0.01$s is simulated. The time interval for beam reconstruction is determined by TIAM to cover the target MU within the HPBW of the beam, which can be calculated using Eq. (\ref{eq16}) and Eq. (\ref{eq17}). In these equations, the parameter $\lambda$ is set as 0.886. When the target MU$_l$ moves rapidly, such as between time index 0 and 100, the beam reconstruction with fixed time interval $\Delta t=0.1s$ cannot adapt to the MU$_l$’s motion, resulting in low beam gain that is insufficient for data transmission. With a fixed time interval, beam reconstruction occurs frequently after time index 100, and the decline in the beam gain is not obvious. For the hybrid codebook beam-tracking algorithm, high beam gain fails to meet transmission requirements when the MU$_l$ moves quickly due to the fixed beam vector. However, the BAB-AR algorithm can reconstruct the beam when the MU$_l$ moves away from the HPBW of the constructed beam. When the movement of MU$_l$ is slow, the beam gain at the MU$_l$ with BAB-AR is superior to the half of the maximum beam gain, and the frequency of beam reconstruction can be reduced with an adaptive beam reconstruction interval calculated by TIAM.
\par The energy efficiency(EE) of the above three mentioned algorithms under different MU movement speed is compared, and the EE at MU$_l$ can be calculated as follows:
\begin{equation}
    \begin{split}
        E{{E}_{l}}=\frac{Rat{{e}_{l}}}{{{p}_{i,l}}+{{N}_{RF}}{{P}_{RF}}+{{N}_{x}}{{N}_{y}}{{P}_{PS}}+{{P}_{BB}}}
    \end{split}
    \label{eq48}
\end{equation}
where $P_{PS}$ is the power consumption of each phase shifter, $P_{RF}$ is the power consumed by each RF chain, $N_{RF}$ is the number of RF chains, and $P_{BB}$ is the power consumption of the baseband. In addition, the rate at MU$_l$ can be calculated as follows:
\begin{equation}
    \begin{split}
        Rat{{e}_{l}}={{\log }_{2}}\left( 1+SN{{R}_{l}} \right)
    \end{split}
    \label{eq49}
\end{equation}
\par
\begin{figure}
\centering
\subfloat[EE at MU$_l$ when MU$_l$ moves fast.]{\includegraphics[width=1.5in,height=1.25in]{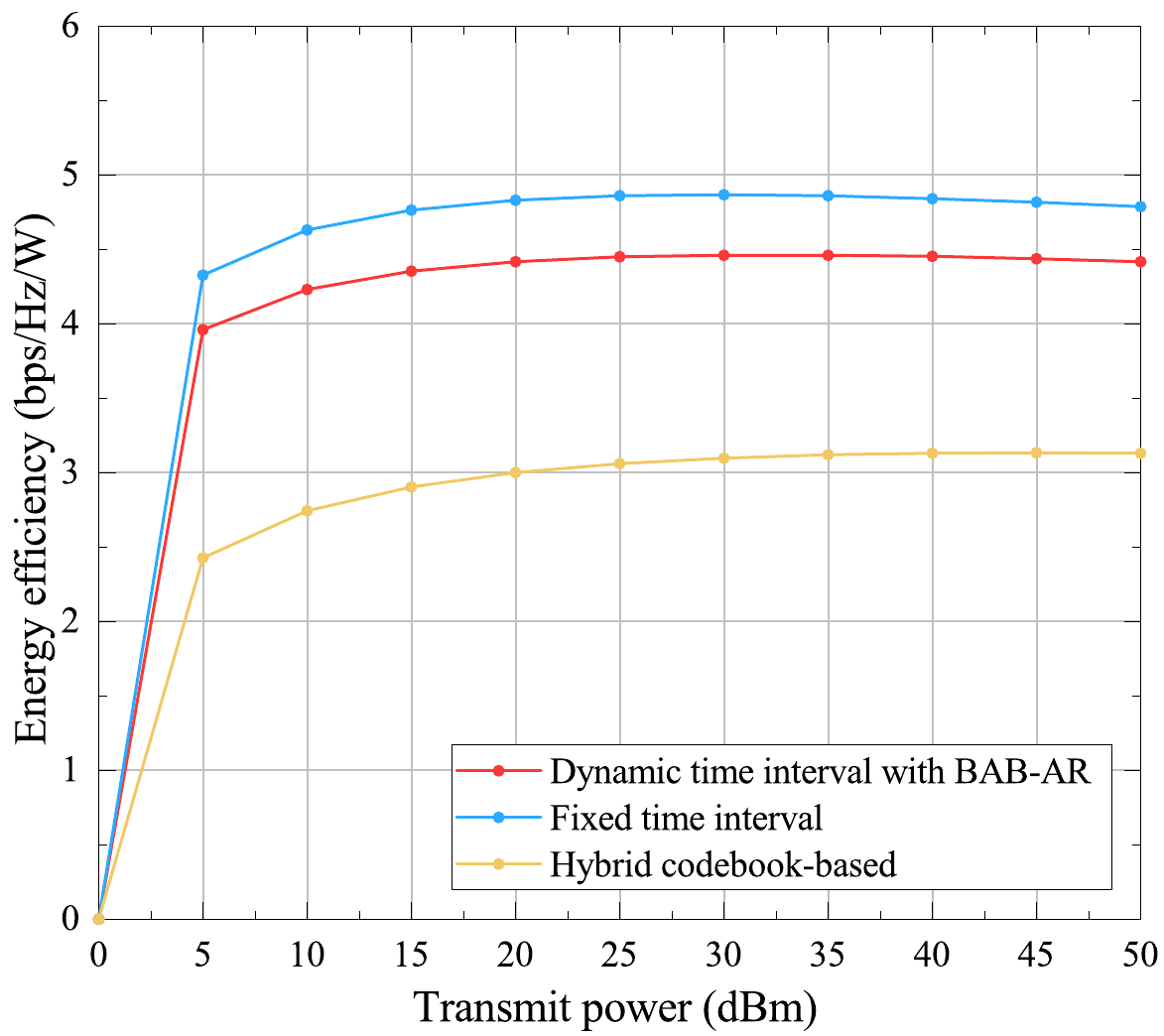}%
\label{fig16a}}
\hfil
\subfloat[EE at MU$_l$ when MU$_l$ moves slowly.]{\includegraphics[width=1.5in,height=1.25in]{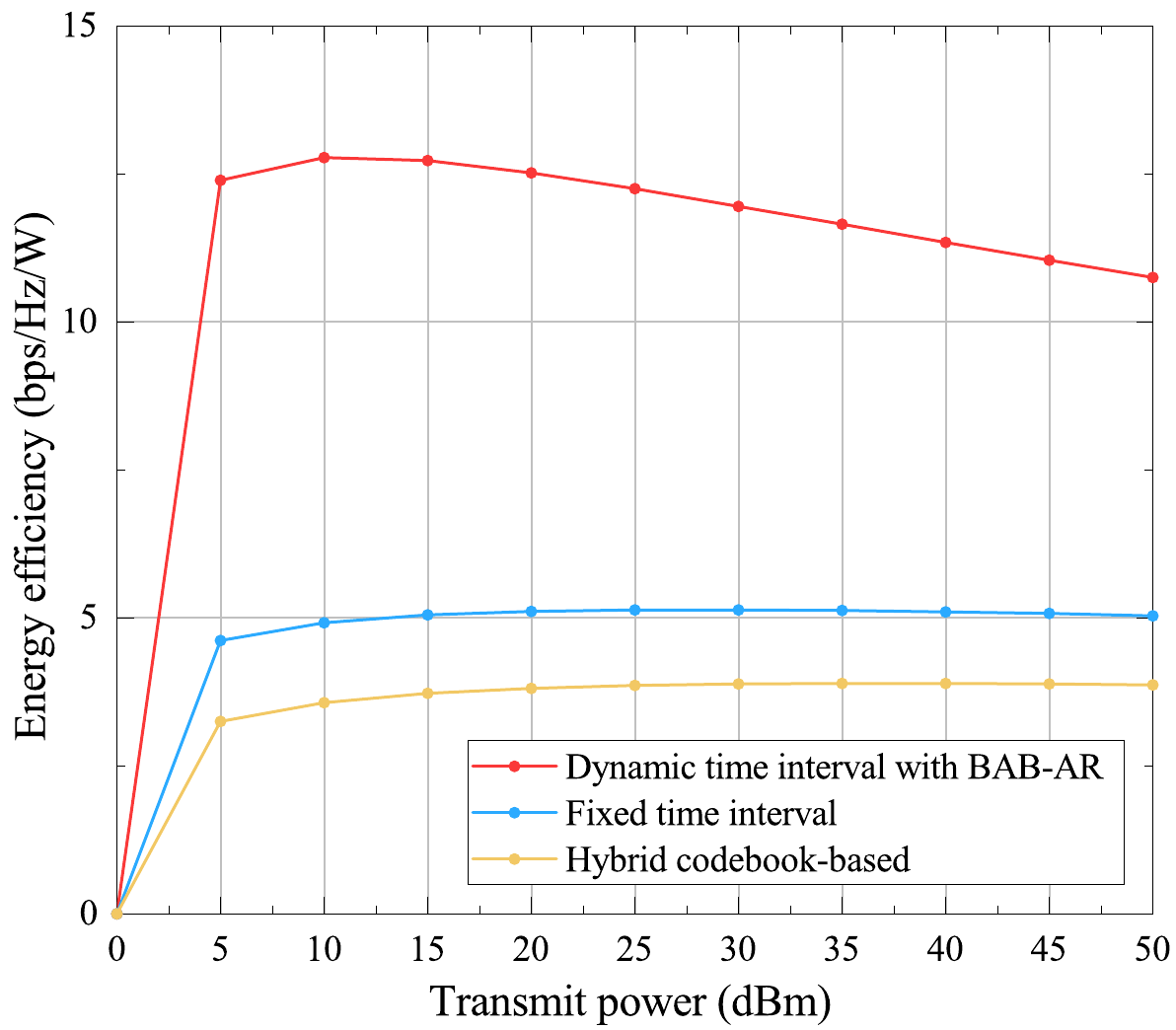}%
\label{fig16b}}
\caption{Energy efficiency (EE) under different transmit power.}
\label{fig16}
\end{figure}

 The growth trend of EE remains stable as the transmit power increases, as shown in Fig. \ref{fig16}. Because the transmit power rises as the impact of the total power consumption increase according Eq. (\ref{eq48}). In this experiment, it is assumed that $P_{PS}$ = 40 mW, $P_{RF}$ = 300 mW, $N_{RF}$ =1, and $P_{BB}$ = 200 mW \cite{ZYXiao2020}. As shown in Fig. \ref{fig16}(a), the EE of BAB-AR is lower compared to the fixed time interval mode when MU$_l$ moves fast, because the frequent beam reconstruction consume much power. However, the beam gain at MU$_l$ under the fixed time interval mode may not meet the required communication quality, as Fig. \ref{fig15} shown. When MU$_l$ moves slowly, as shown in Fig. \ref{fig16}(b), the EE of BAB-AR at MU$_l$ improves significantly by up to 136.46\% and 215.99\% compared to that of the fixed time interval mode and hybrid-codebook algorithm, respectively. Because BAB-AR algorithm not only ensures that the beam gain remains at half of the maximum beam gain that the antenna can provide, but also reduces the number of beam reconstruction when the target MU$_l$ moves slowly.
\par
\begin{figure}
\centering
\subfloat[Average transmit rate at MU$_l$ when MU$_l$ moves fast.]{\includegraphics[width=1.5in,height=1.25in]{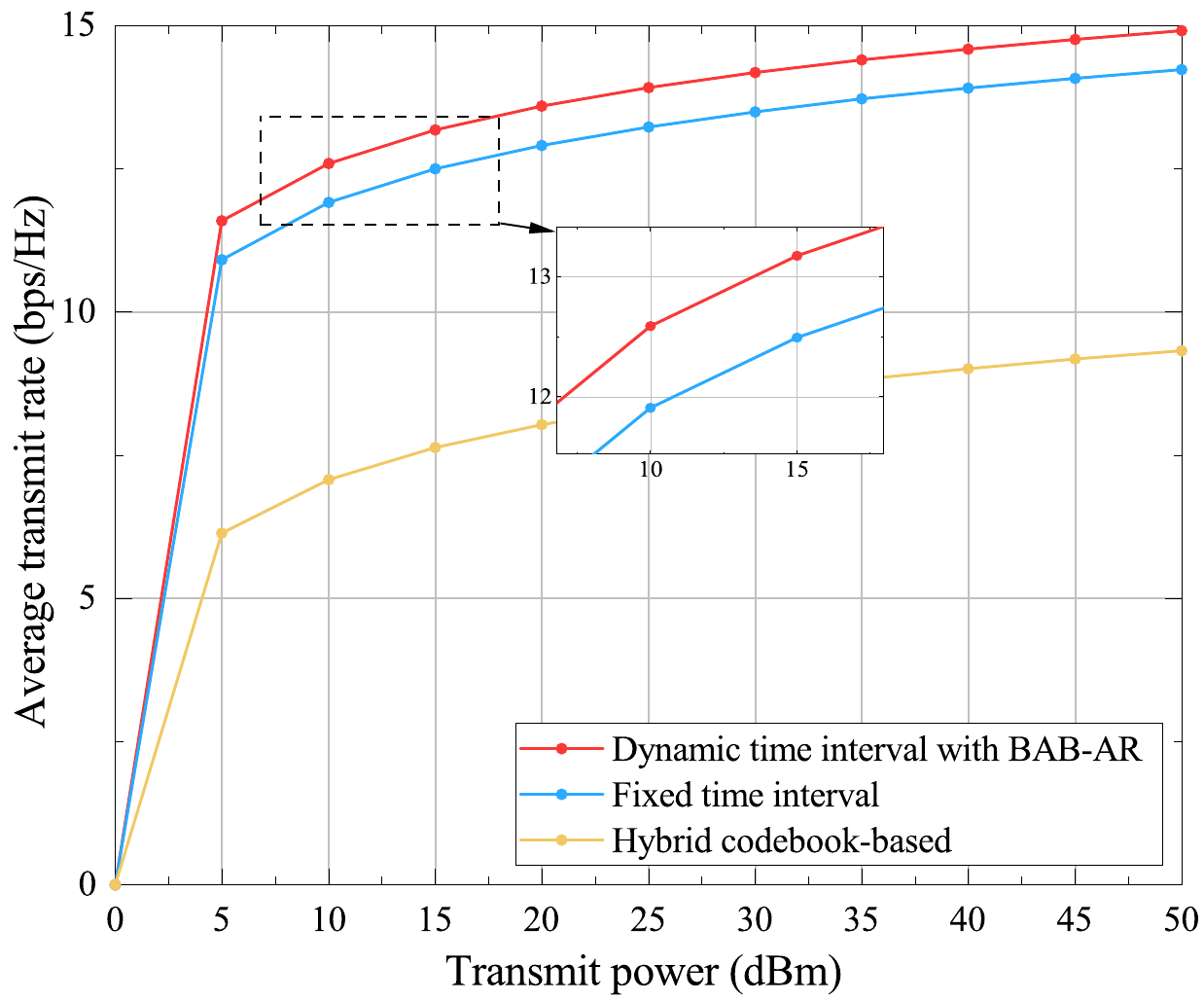}%
\label{fig17a}}
\hfil
\subfloat[Average transmit rate at MU$_l$ when MU$_l$ move slowly.]{\includegraphics[width=1.5in,height=1.25in]{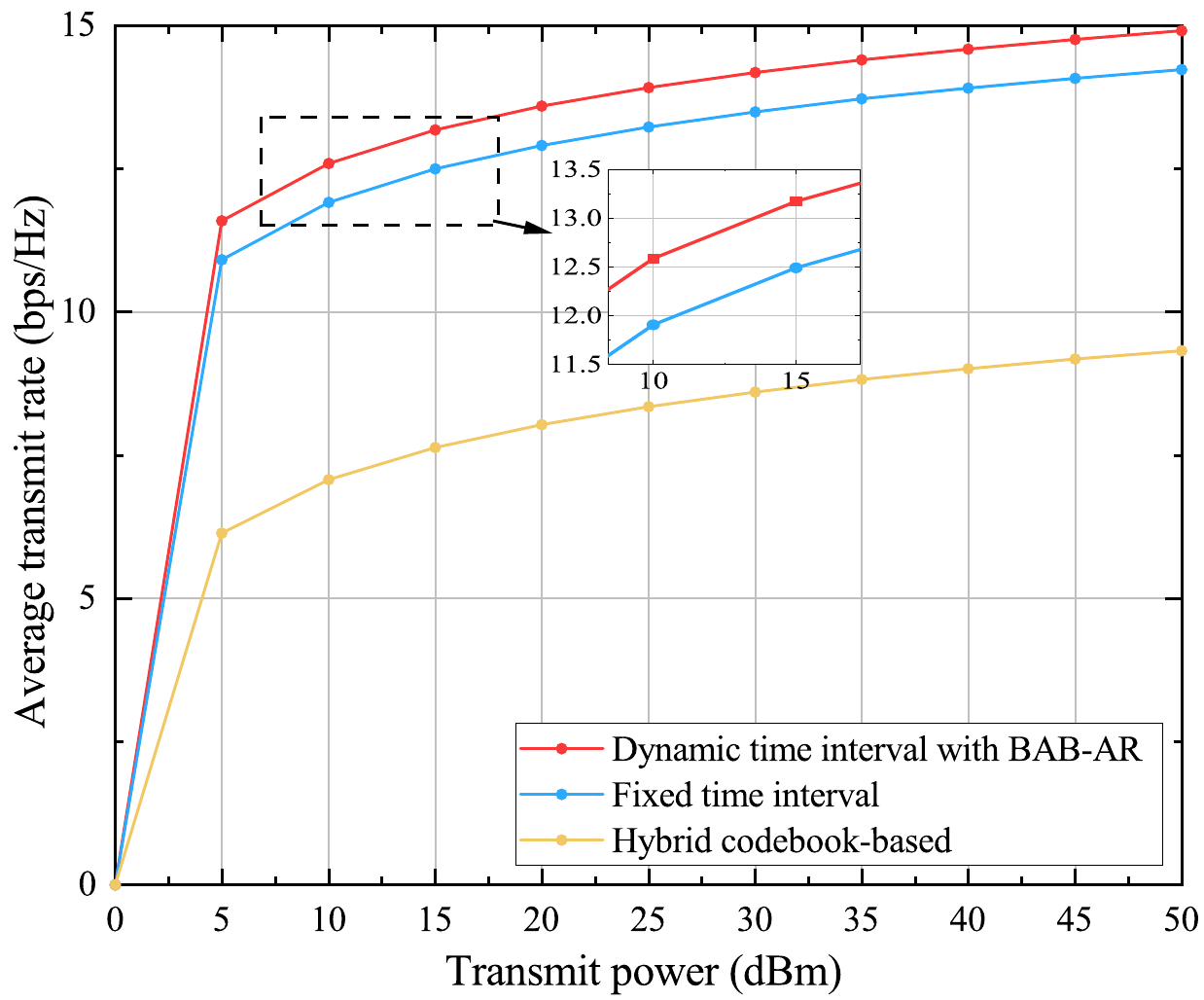}%
\label{fig17b}}
\caption{Average transmit rate under different transmit power.}
\label{fig17}
\end{figure}

 The average transmit rate at MU$_l$ is evaluated under different transmit power level and speed, and the noise power is set as -100 dBm. The hybrid codebook's average transmit rate is not high enough due to the low beam gain at MU$_l$. In addition, when MU$_l$ moves fast as shown in Fig. \ref{fig17}(a), the average transmit rate of BAB-AR is improved by up to 5.21\% and 67.63\% compared to that of the fixed time interval mode and hybrid-codebook algorithm. Because the beam in BAB-AR can be reconstructed according to the movement of the MU$_l$ and cover the MU$_l$ with HPBW. When MU$_l$ moves slowly as shown in Fig. \ref{fig17}(b), the BAB-AR reconstructs beam only once, but the fixed time interval mode reconstructs beam 10 times. However, the difference in average transmit rate between BAB-AR and fixed time interval mode is less than 0.5. 
\subsection{Limitation of the BAB-AR algorithm}
\begin{figure}
    \centering    \includegraphics[width=2.5in,height=1.43in]{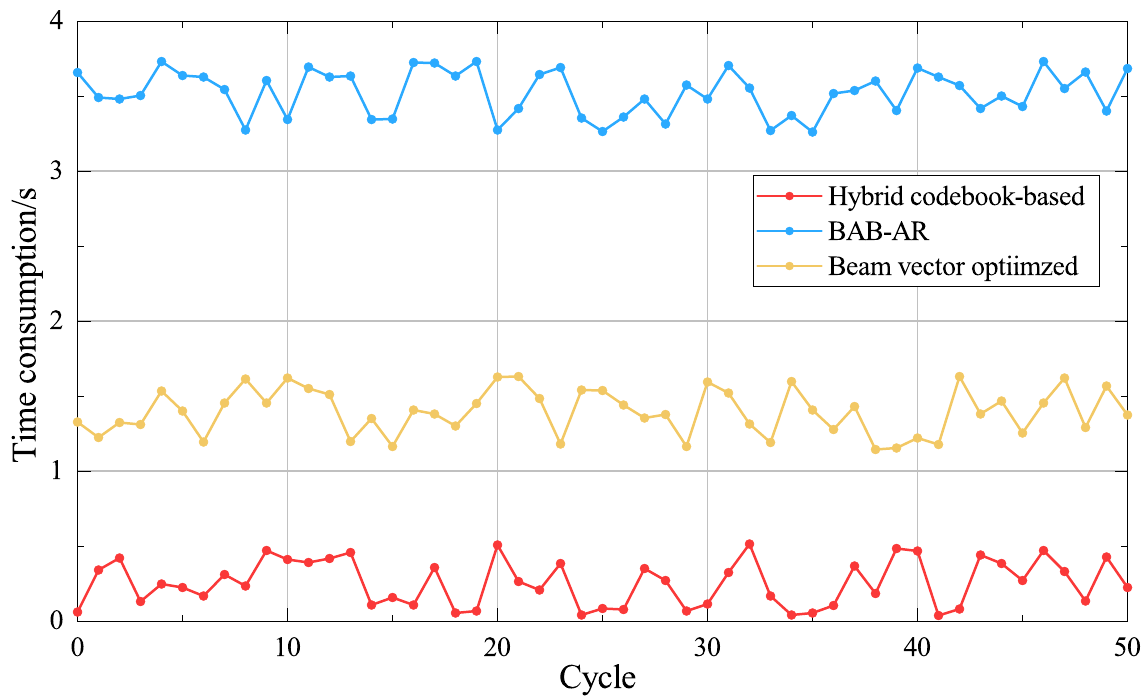}
    \caption{Running time of different algorithms.}
    \label{fig18}
\end{figure}

The time for constructing the optimal beams has a significant impact on the effectiveness of the algorithm in real scenarios. In particular, the shorter the time consumed, the more beneficial for real-time beam alignment in high-mobility scenarios. The complexity of the codebook beam-tracking algorithm \cite{AAlkhateeb2014} is $\mathcal{O}$($N_c$), where $N_c$ is the number of designed codebook, while the complexity of the beam vector-optimized algorithm \cite{XChen2020} and the proposed BAB-AR are $\mathcal{O}$($N_{iter} \times N_{3.5}$) and $\mathcal{O}$($N_{ULA} \times N_{iter} \times N_{pop}$), respectively, as mentioned in Section IV. In Fig. \ref{fig18}, we set the number of antennas as $16 \times 16$, and compare the time consumption of BAB-AR algorithm, beam vector-optimized beam-tracking algorithm \cite{XChen2020}, and hybrid codebook-based algorithm \cite{AAlkhateeb2014} in each cycle. BAB-AR consumes more time than that of the codebook-based beam alignment algorithm. However, compared to the algorithm proposed in \cite{XChen2020}, the proposed BAB-AR algorithm can reduce the time consumption by 44.89\%. 
\par
\begin{figure}
    \centering   \includegraphics[width=2.5in,height=1.43in]{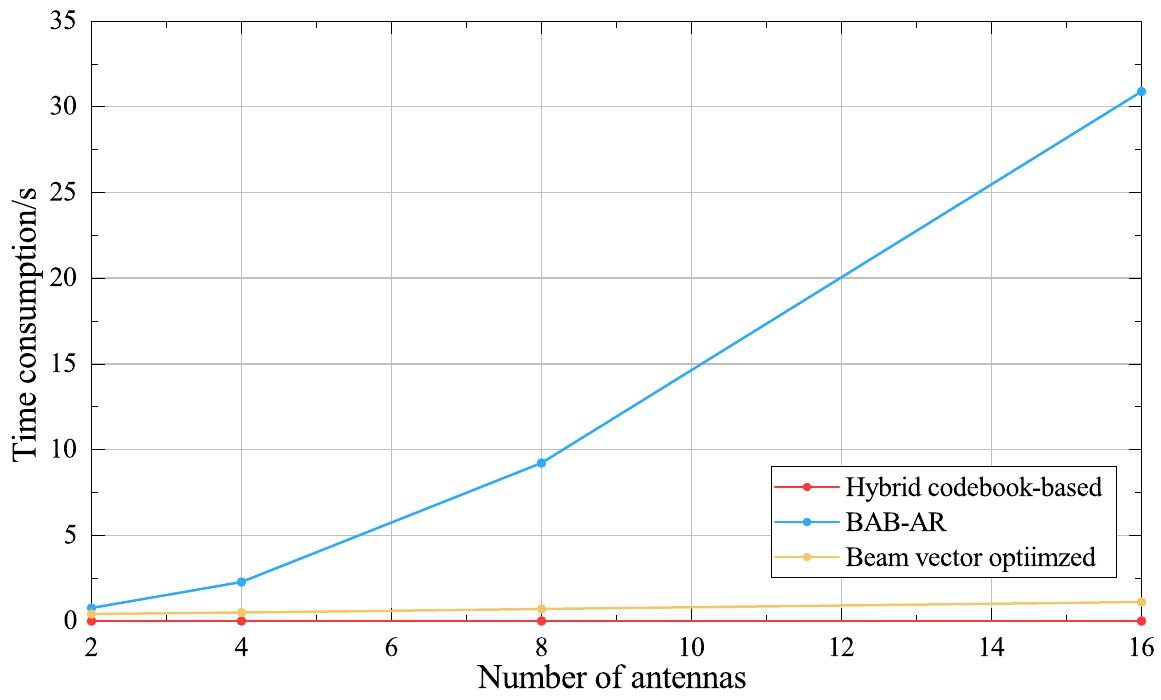}
    \caption{Time consumption under different numbers of antennas.}
    \label{fig19}
\end{figure}

 The impact of the numbers of UPA's antennas on the time consumption of different algorithms is shown in Fig. \ref{fig19}. As the number of antennas increase, so does the time required to obtain the optimal beam vector. The performance of BAB-AR is better than that of the beam vector optimized \cite{XChen2020}, and its time consumption does not increase obviously with the number of antennas. However, BAB-AR still has a higher time consumption than that of the hybrid codebook-based algorithm. Despite reducing the dimension of the optimized problem, BAB-AR's time consumption is proportional to the number of ULA antennas equipped on A-UAV, the number of iterations and the initial population, resulting in extra time for beam selection compared to that of the hybrid-codebook algorithm.
\par
\begin{figure}
\centering
\subfloat[Trajectory of MU with different speed.]{\includegraphics[width=1.5in,height=1.25in]{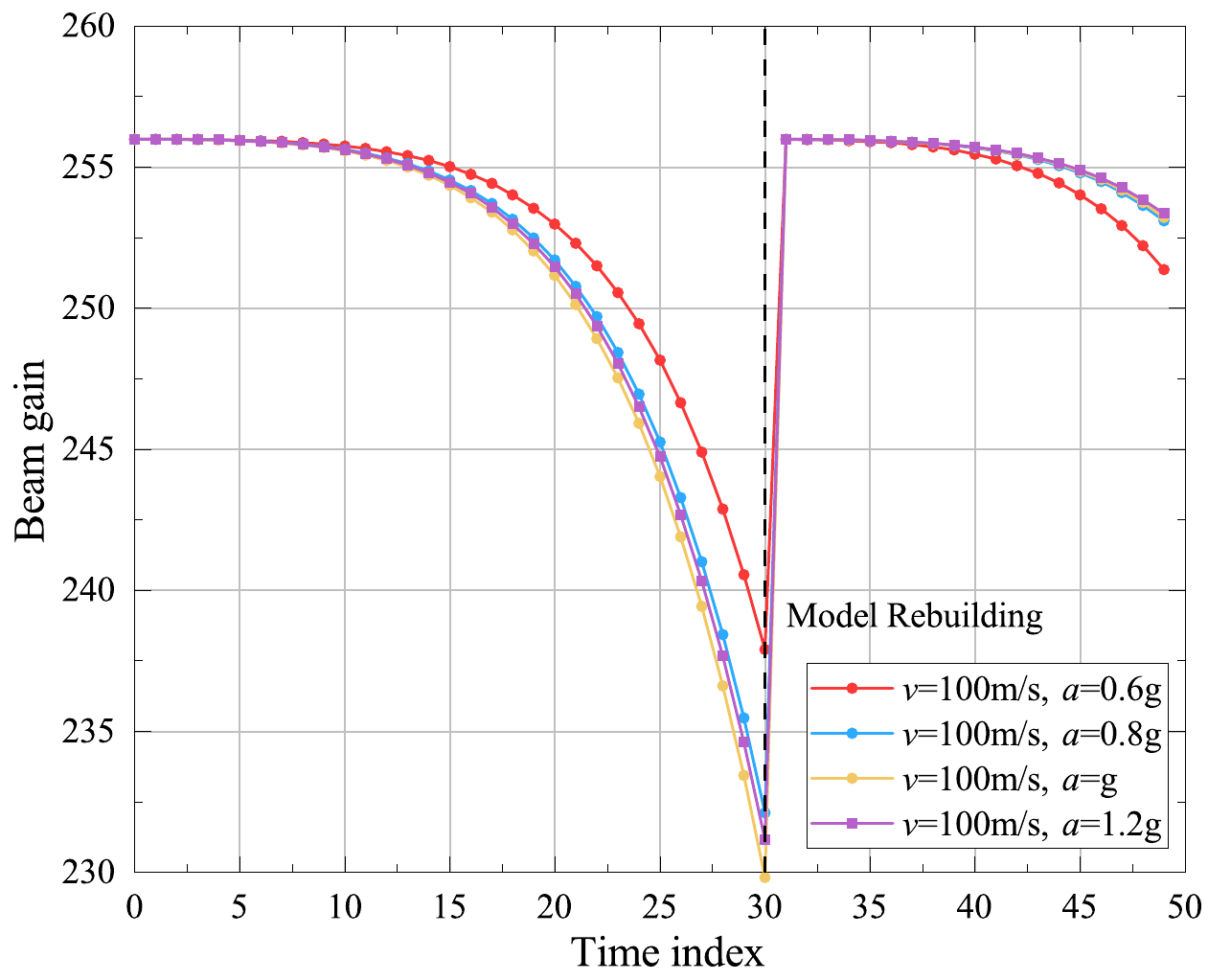}%
\label{fig20a}}
\hfil
\subfloat[Fixed trajectory of MU.]{\includegraphics[width=1.5in,height=1.25in]{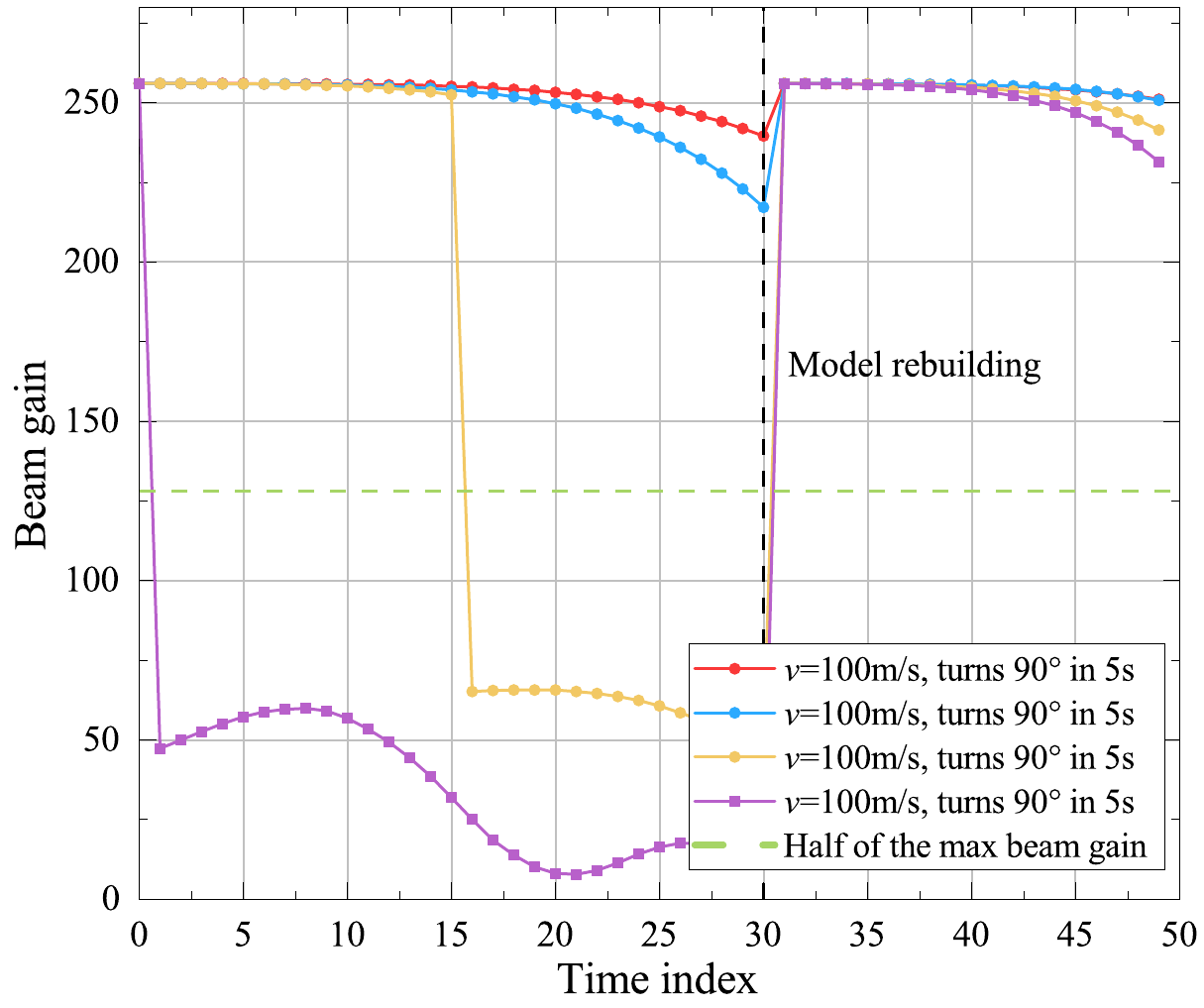}%
\label{fig20b}}
\caption{Beam gain received at the MU$_l$ with different algorithms.}
\label{fig20}
\end{figure}

 We also evaluated the BAB-AR algorithm's applicability for different trajectories with an initial velocity of $v$=100 m/s. As shown in Fig. \ref{fig20}(a), acceleration of speed does not affect the accuracy of the BAB-AR algorithm, but the curvature of the trajectory can reduce the accuracy of position prediction. In Fig. \ref{fig20}(b), when the target MU$_l$ turns 160° in 5s, the trajectory from $t_n$=16 to $t_n$=20 is unrelated to the historical trajectory, and the beam gain of BAB-AR improves at $t_n$=30. However, When the MU$_l$ turns 180° in 5s, the time required for BAB-AR to reconstruct beam cannot adapt to changes in trajectory. Because the trajectory from $t_n$=0 to $t_n$=30 has little correlation with previous trajectories. Therefore, the beam gain at the MU$_l$ cannot exceed half of the maximum beam gain. In summary, the BAB-AR algorithm cannot adapt to MUs with too little trajectory information to predict their trajectories accurately.
\section{Conclusion}
This study proposes an accurate beam-tracking algorithm for MUs. UAV-BSs consisting of A-UAVs and U-UAVs are employed \sethlcolor{-blue}to provide beam tracking for the MUs with adaptive beam reconstruction. U-UAVs without GPS can be located with high accuracy by A-UAVs using the proposed GDCSA algorithm. Furthermore, a TIAM is proposed for the adaptive reconstruction of the beam, and MUs can be covered with HPBW. Simulation results show that the predictive angle relative error of the GDCSA algorithm is within 0.2 under different motion patterns, which realizes a higher SNR and transmit rate at U-UAV compared to that of the angle-aware UAV beam tracking with machine model. Meanwhile, when the target MU moves fast, BAB-AR can maintains the value of beam gain at the target MU as half of the maximum beam gain that the antenna can provide. When the target MU moves slowly, the energy-efficiency at MU can be improved by up to 136.46\% and 215.99\% compared to that of the fixed time interval mode and the codebook-based algorithm, respectively. However, BAB-AR maynot perform effectively when the MU trajectory has a varying radian, such as the S-bend trajectory. Therefore, furture work will focus on modeling MU trajectories to improve the accuracy of beam alignment.
\bibliography{refs.bib}
\bibliographystyle{IEEEtran}
\vspace{8pt}
\begin{IEEEbiography}[{\includegraphics[width=1in,height=1.25in,clip,keepaspectratio]{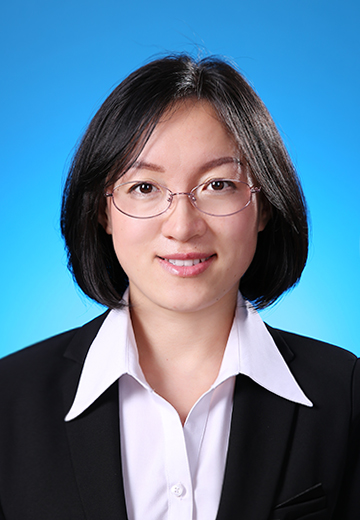}}]{Jing Zhang (Member IEEE)} received the Ph.D degree in computer science and technology from Jilin University, Changchun, China, in 2015. She is Postdoctoral fellow at the College of Communication Engineering, Jilin University. Currently, she is an Associate Professor with the College of Computer Science and Technology, Changchun University of Science and Technology. Her research interests include internet of Things, blockchain, and complex network.
\end{IEEEbiography}
\vspace{-4\baselineskip}

\begin{IEEEbiography}[{\includegraphics[width=1in,height=1.25in,clip,keepaspectratio]{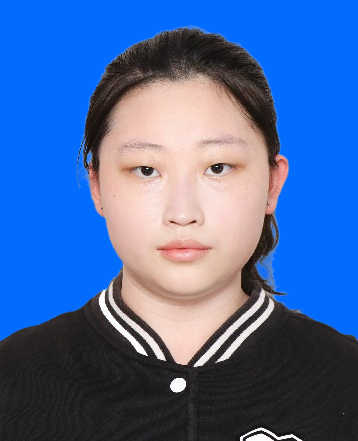}}]{Sheng Gao} received the B.Eng. degree in computer science and technology from Changchun University of Science and Technology, in 2021. And she is working towards the M.E. degree in Changchun University of Science and Technology from September 2021. Her research focuses on beam tracking.
\end{IEEEbiography}

\vspace{-4\baselineskip}
\begin{IEEEbiography}[{\includegraphics[width=1in,height=1.25in,clip,keepaspectratio]{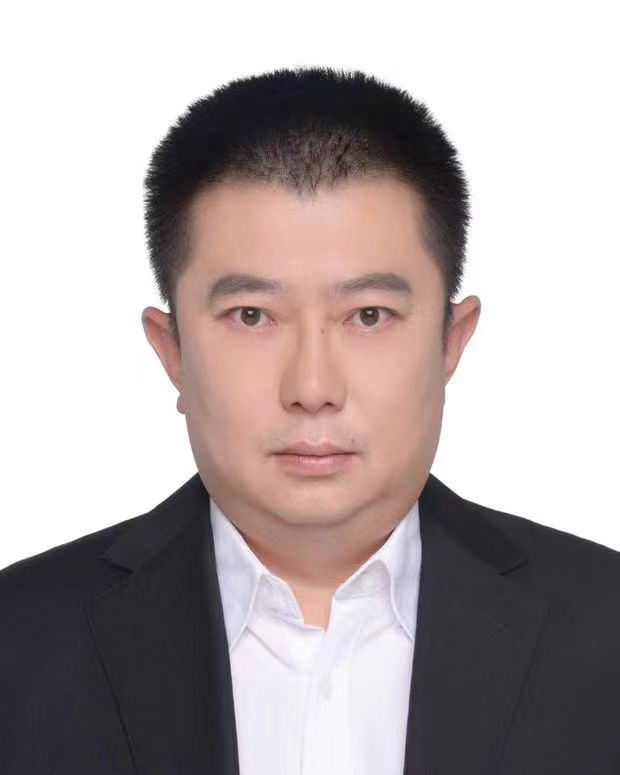}}]{Xin Feng} received the Ph.D degree in computer science and technology from Changchun University of Science and Technology in 2014. He is currently a research fellow with the College of Computer Science and Technology, Changchun University of Science and Technology. His research interests include clustered storage and parallel processing, expert systems and intelligent diagnostics.
\end{IEEEbiography}

\vspace{-4\baselineskip}
\begin{IEEEbiography}[{\includegraphics[width=1in,height=1.25in,clip,keepaspectratio]{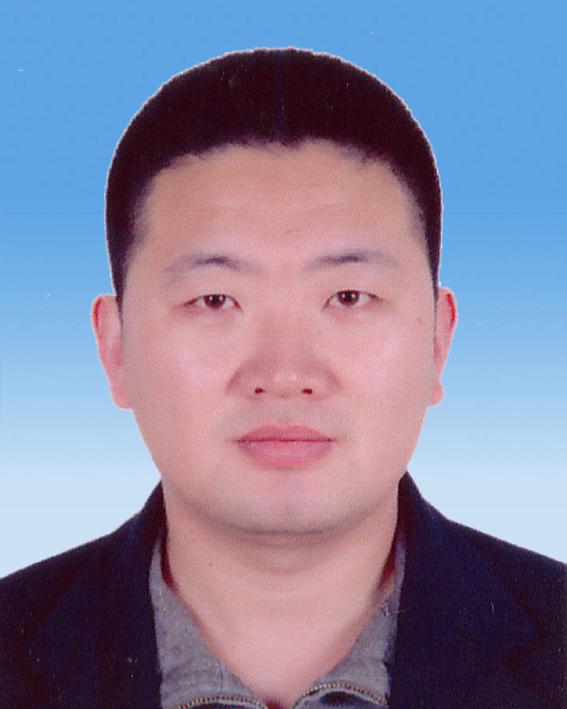}}]{Hongwei Yang} received the Ph.D degree in computer science and technology from Changchun University of Science and Technology in 2021. He is currently a Professor with the College of Computer Science and Technology, Changchun University of Science and Technology. His research interests include internet of Things, software engineering and information systems, and data Mining.
\end{IEEEbiography}

\vspace{-4\baselineskip}
\begin{IEEEbiography}[{\includegraphics[width=1in,height=1.25in,clip,keepaspectratio]{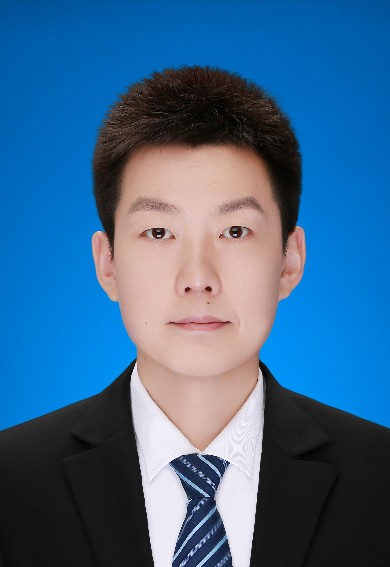}}]{Geng Sun (Member IEEE)} received the B.S. degree in communication engineering from Dalian Polytechnic University in 2011 and the Ph.D. degree in computer science and technology from Jilin University in 2018. He was a Visiting Researcher with the School of Electrical and Computer Engineering, Georgia Institute of Technology, USA. He is currently an Associate Professor with the College of Computer Science and Technology, Jilin University. His research interests include wireless networks, UAV communications, collaborative beamforming, and optimizations.
\end{IEEEbiography}

\end{document}